\pdfoutput=1
\documentclass[12pt,a4paper]{article}
\usepackage[margin=2cm]{geometry}
\usepackage{psfrag}
\usepackage{amsmath}
\usepackage{amsfonts}
\usepackage{amssymb}
\usepackage{graphicx}
\usepackage{pstricks}
\usepackage{hyperref}
\usepackage[utf8]{inputenc}
\usepackage[T1]{fontenc}
\usepackage{lmodern}
\usepackage{lscape,epsf}

\DeclareMathOperator{\miso}{mini-ISO}

\usepackage{graphicx} 
\usepackage{cancel}

\newcommand{\ignore}[1]{}
\newcommand{\be}{\begin{equation}} \newcommand{\ee}{\end{equation}}
\newcommand{\ba}{\begin{eqnarray}} \newcommand{\ea}{\end{eqnarray}}
 \renewcommand{\bf}{\textbf}

 \renewcommand{\i}{\mathrm{i}}

\newcommand{\ADTT}{ATLAS-$d_{12}$ tagger}

\newcommand{\GeV}{{\rm\ GeV}}

\newcommand{\TeV}{{\rm\ TeV}}

\newcommand{\mch}{MCH4$_5$~}

\newcommand{\stp}{$\rm singleTP_{narrow}$}
\newcommand{\notop}{$\rm lightTP_{notop}$}

\begin{document}

\begin{titlepage}

\begin{flushright}
{CAFPE-185/14 \\ CERN-PH-TH-2014-216
\\ DESY 14-213 \\ SISSA 60/2014/FISI\\
UG-FT-315/14 }
\end{flushright}

\begin{center}
{\Huge \bf {The Elusive Gluon}
}
\vskip 1.2cm
  {   Mikael Chala$^{1,2}$, Jos\'{e} Juknevich$^{3,4}$, Gilad
    Perez$^{5}$\\ and Jos\'{e} Santiago$^{1,6}$} 
\vskip 0.4cm

{\it
$^{1}$CAFPE and Departamento de F\'{\i}sica Te\'{o}rica y del Cosmos, \\ Universidad de Granada, E-18071 Granada, Spain\\
$^{2}$ DESY, Notkestrasse 85, 22607 Hamburg, Germany \\
$^{3}$SISSA/ISAS, I-34136 Trieste, Italy\\
$^{4}$INFN - Sezione di Trieste, 34151 Trieste, Italy\\
$^{5}$Department of Particle Physics and Astrophysics, \\ Weizmann Institute of Science, Rehovot 76100, Israel\\
$^{6}$CERN, Theory Division, CH1211 Geneva 23, Switzerland\\

}

\end{center}

\vspace{0.5cm}
\begin{abstract}
We study the phenomenology of vector resonances in the context of natural
composite Higgs models.  
A mild hierarchy between the fermionic partners and the
vector resonances can be expected in these models based on the
following arguments.
Both direct and indirect (electroweak and flavor precision) constraints
on fermionic partners are milder than the ones on spin one
resonances. 
Also the naturalness pressure coming from the top partners is stronger
than that induced by the gauge partners. This observation implies that
the search strategy for vector resonances at the LHC needs to be modified. 
In particular, we point out the importance of heavy gluon decays (or
other vector resonances) to top partner pairs that were overlooked in
previous experimental searches at the LHC.  These searches focused on
simplified benchmark models in which the only new particle beyond the
Standard Model was the heavy gluon. It turns out that, when
kinematically allowed, such heavy-heavy decays make the heavy gluon
elusive, and the bounds on its mass can be up to 2 TeV milder than in
the simpler models considered so far for the LHC14. 
We discuss the origin of this
difference and prospects for dedicated searches. 
\end{abstract}

\end{titlepage}

\tableofcontents

\section{Introduction}\label{sec:intro}

The LHC discovery of the Higgs boson, with a light mass and couplings
consistent with the Standard Model (SM) prediction, reinforces the
fine tuning problem. In the SM the Higgs mass is radiatively
unstable and it has uncontrolled sensitivity to microscopic
dynamics. 
A simple possibility to stabilize the Higgs mass and the electroweak
scale in a controlled manner is to add new fields to the SM, with the same
gauge quantum numbers as the SM fields, such that the contributions of
the new fields to the Higgs mass eliminate the UV sensitivity. 
The most severe known sensitivity of the Higgs to quantum corrections
arises as a result of its large coupling to the top quark. To ensure the
stabilization of the electroweak scale, the virtual contributions of
some of the new particles to the Higgs mass should cancel the
contributions coming from the SM top quarks. These new states are
collectively denoted as top partners. In known examples, the partners
might be scalars as in the case of supersymmetry or fermions as in the
case of composite Higgs models (CHMs). Naturalness also requires the
presence of additional states, partners of the electroweak gauge
bosons, fermions in the supersymmetric case and massive vectors in
CHMs.  

Our focus in this paper is to study the interplay between the
collider phenomenology of the massive vectors and the top partners
within a class of natural CHMs.  
As we shall see below, the search strategy that was chosen so far by
the LHC experiments regarding the vector resonance might be incomplete
and can potentially be improved in an essential manner. 
Our claim stems from the following three observations regarding the
status of natural composite Higgs models (see also~\cite{PerezTalk}): 
\begin{itemize}
\item The direct constraints on the mass of the top partners are
  weaker than those of the vector resonances. The lower bound on the
  mass of composite-Higgs fermionic states is roughly 800\,GeV (see {\it
    e.g.} Refs.~\cite{Chatrchyan:2013uxa,Aad:2014efa} for
  recent results), while the corresponding lower bounds on the mass of
  a colour-octet spin-one resonance mass is in the 2-2.5\,TeV region  (see {\it e.g.}
  Refs.~\cite{TheATLAScollaboration:2013kha,Chatrchyan:2013lca}).  
\item The indirect constraints on the mass of the top partners
  are weaker than those of the vector resonances. The lower bounds from
  electroweak (EW) precision tests~(see~\cite{Carena:2007ua,Anastasiou:2009rv,Grojean:2013qca}) and flavor
  physics~\cite{BiancofioreDaRoldJagerGPWeilertoappear} on the fermion
  partners are roughly of a TeV, while the corresponding lower bounds
  on the vector resonance masses is in the multi-TeV range. 
\item The naturalness pressure on the top partners is stronger
  than that of the vector resonances, as is well known (see {\it
    e.g.}~\cite{Schmaltz:2002wx} and Refs. therein). In a natural
  theory the top partners are required to lie below the TeV scale whereas the
  vector resonances can have masses beyond the TeV scale.  
  Furthermore, the combination of LEP and Tevatron data constrains
  the model's decay constant $f$ to lie above the
  $f>{\cal O} (800\,\rm GeV)$
  scale~\cite{Grojean:2013qca,Ciuchini:2013pca}. Therefore the
  composite fermion resonances would be somewhat heavier with masses
  probably larger than $f\,.$ Thus, requiring similar level of tuning
  on the vector resonances will send their masses to the multi-TeV
  range.  The measured value of the Higgs mass further increases this
  pressure as pointed out
    in~\cite{Matsedonskyi:2012ym,Redi:2012ha,Marzocca:2012zn,Pomarol:2012qf,Panico:2012uw,Pappadopulo:2013vca} 
    (see \cite{Barnard:2013hka,Carmona:2014iwa} for a discussion of
    effects that could partially alleviate this pressure). 
\end{itemize}
Adding the information in these items leads to a potentially viable
spectrum of natural pseudo Nambu-Goldstone Boson (pNGB) 
composite models where the top partners are
relatively light, with masses around the TeV scale, while the spin one states
such as the $Z'/W'$ and the colour octet resonances or the
Kaluza-Klein (KK) excitations in warped extra-dimensional
Randall-Sundrum (RS) models~\cite{Randall:1999ee, ArkaniHamed:2000ds} are expected to have
masses in the multi-TeV region.  
This mild hierarchy between the masses of the vector resonances and
the fermionic partners not only improves the consistency
of the framework but it also suggests a qualitative change in the
current search strategies for 
the composite vector resonances. The reason is simple: in the
above set up we generically expect that the inter-composite couplings,
or the couplings between the resonances, would dominate over the
couplings between the resonance and the SM fields (that cannot be all
composite due to various constraints).    
Thus, unlike in the original theoretical constructions, the mild
hierarchy in scales implies that the resonances such as the KK
gluon~\cite{Agashe:2006hk,Lillie:2007yh} and their EW
counterparts~\cite{Agashe:2007ki} (including the celebrated $Z'$)
would preferably decay to pairs of top partners instead of pairs of SM
fields, such as tops, the final state that has been most frequently
analysed thus far.

We study in this paper in a quantitative way the implications that top
partners have on heavy gluon searches (see~\cite{Vignaroli:2014bpa}
for a related analysis of $W^\prime$ resonances). 
The complementary effect, namely the implications that heavy gluon
resonances have on top partner searches was discussed in detail in the
context of holographic composite Higgs models in~\cite{Carena:2007tn}.
We will show that the larger
width and the large number of new channels, typically involving
$t\bar{t}+X$ in the final state with $X$ a pair of SM gauge or Higgs
bosons, make the heavy gluon much more elusive than the
one in which top 
partners are absent. 
Current bounds can be easily a few hundred GeV less
stringent for realistic heavy gluons than the ones currently being
reported. At the LHC14 this difference can go up to almost $2$
TeV (see also~\cite{SantiagoTalk}). We discuss the interplay of
resolved and boosted analyses in 
these searches and we finally propose simple extensions of current
analyses that would allow one to recover a good fraction of the lost
sensitivity. The emphasis will be in the natural region of parameter
space in which pair production of top partners is kinematically
open. The case of heavier top partners for which single production in
association with a SM (top or bottom) quark is the dominant channel
has been studied in detail
in~\cite{Barcelo:2011vk,Barcelo:2011wu,Bini:2011zb,Carmona:2012jk,Chala:2013ega}
and will be used here just for
comparison purposes.

The rest of the article is organized as follows. In Section \ref{sec:model} we
describe the main features of our minimal CHM.  In
Section~\ref{sec:signatures} we discuss the main implications of top
partners on the phenomenology of the heavy gluon. In
Sections~\ref{sec:exp_limits} and \ref{sec:analysis} we summarize the existing
experimental bounds on heavy gluons and describe in detail our reconstruction
method. Our main results, the current and predicted bounds for the
LHC14 on the heavy gluon in the presence of top partners are reported
in Section~\ref{sec:results}, where we also present our proposed
search strategies to improve the sensitivity to the heavy gluon.
Finally we conclude in Section~\ref{sec:conclusions}.
A detailed description of the model is provided in the Appendix.

\section{The model} \label{sec:model}

The general discussion in the previous Section shows that, quite
generically, the phenomenology of the heavy gluon in realistic
CHMs is likely to be very different from the one in
the models currently being used to interpret experimental searches. In order
to make a quantitative estimate of the effects of such differences and
their implications for LHC searches,
we consider the Minimal Composite Higgs model, based on the
$SO(5)/SO(4)$ coset, with composite fermions transforming in the
vector representation \textbf{(5)} of
$SO(5)$~\cite{Agashe:2004rs,Contino:2006qr}. This is the minimal model
that contains only the Higgs doublet as the pNGB of the symmetry
breaking and incorporates custodial symmetry to protect the $T$
parameter and the $Z b_L \bar{b}_L$
coupling~\cite{Agashe:2006at}. For the sake of simplicity
we are going to consider a
simplified version of this model, developed in~\cite{DeSimone:2012fs},
in which the right-handed (RH) top quark is fully composite and
only the first level of fermion resonances is included. The model is
denoted MCH4$_5$, where the $5$ indicates the $SO(5)$
representation of the composite operator that mixes with the SM
left-handed quark doublets realizing the partial compositeness
scenario, and the $4$ stands for the $SO(4)$ representation
of the lightest fermion resonances which read explicitly
\begin{equation}\Psi = \frac{1}{\sqrt{2}} \left( \begin{array}{c}
iB- i X_{5/3}\\
B+ X_{5/3}\\
iT+i X_{2/3}\\
-T + X_{2/3}
\end{array} \right).
\end{equation}
In terms of $SU(2)_L\times U(1)_Y$ representations, the fourplet
$\Psi$ gives rise to two doublets. One doublet $(T,B)$ with
hypercharge 1/6, as the SM left-handed doublet, and a second doublet
$(X_{5/3}, X_{2/3})$ with hypercharge 7/6, containing an exotic state
with charge 5/3, $X_{5/3}$, and a charge 2/3 state, $X_{2/3}$.
After electroweak symmetry breaking there is a 
linear combination of the two charge $2/3$ quarks, denoted
$X^\prime_{2/3}$, that is degenerate with $X_{5/3}$ whereas the orthogonal
combination, denoted $T^\prime$, and $B$ are somewhat heavier and with
a small mass splitting between them. They decay almost in all the
parameter space into a top quark and a SM gauge or Higgs boson, with
approximately equal branching ratio (\textit{BR}) into all open channels
\begin{eqnarray}
&&BR(X_{5/3}\to t W^+)
=
BR(B\to t W^-)=1, \\
&&BR(X^\prime_{2/3}\to t Z)
\approx
BR(X^\prime_{2/3}\to t H)
\approx
BR(T^\prime \to t Z)
\approx
BR(T^\prime \to t H)
\approx \frac{1}{2}.
\end{eqnarray}

Apart from the scale characterizing the strong coupling, that we
fix to $f=800$ GeV, and
the mass of the degenerate fermion resonances,
$M_\Psi=M_{X^\prime_{2/3}}=M_{X_{5/3}}$, there are only three order
one dimensionless parameters in the original model. One of these
parameters is fixed by the top mass (we take it to be $y$ in the
notation of Ref.~\cite{DeSimone:2012fs}, see the Appendix for details) 
and the other two have a small effect on the
phenomenology that we are investigating so we just fix them to
$c_1=0.7$ and  $c_2=1.7$,
again in the notation of Ref.~\cite{DeSimone:2012fs}.

Regarding the heavy gluon we asume that there is a composite heavy
vector color octet that couples to the composite quarks (including
$t_R$) with a
coupling $g_c$ and an elementary massless color octet that couples
with the elementary fields with a coupling $g_e$. The two color-octet
vectors mix linearly in such a way that a linear combination remains
massless, the partially composite SM gluon, with a coupling
\begin{equation}
g_s=g_e \cos \theta_3=g_c \sin \theta_3.
\end{equation}
The orthogonal combination is the heavy gluon we are interested
in. The coupling of elementary and composite fermions to the heavy
gluon read
\begin{eqnarray}
G \bar{\psi}_{\mathrm{elem}} \psi_{\mathrm{elem}} &:&
-\frac{g_s^2}{\sqrt{g_c^2 - g_s^2}}, \\
G \bar{\psi}_{\mathrm{comp}} \psi_{\mathrm{comp}} &:&
\sqrt{g_c^2 - g_s^2}.
\end{eqnarray}
Once we go to the physical basis for the fermions, they become
partially composite and their couplings depend on the degree of
compositeness. For simplicity we consider that all first two
generation quarks, together with the RH component of
the bottom quark, are purely elementary.  The heavy gluon brings two
new parameters in the game, the composite coupling $g_c$ and the heavy
gluon mass $M_G$. The mass will be taken as a free parameter that we
scan over while we fix the composite coupling to $g_c=4$ in our analyses. This
is a somewhat smaller value of the one that would correspond to the
original RS model, which would lead to too large a heavy gluon width
when decays into top partners are kinematically open. In practice this
means that the coupling of the heavy gluon to mostly composite states
is smaller and the one to mostly elementary fields is larger than in
the original RS model.

To summarize, we fix the following values of the parameters:
\begin{equation}
g_c=4,\qquad f=800\mbox{ GeV}, \qquad c_1=0.7, \qquad c_2=1.7~.
\end{equation}
This choice of $g_c$ impies a coupling of the heavy gluon to the light
SM quarks  $g_{Gqq}=-0.377$.
We take two benchmark values for the top partner mass parameter $M_\Psi$:
\begin{eqnarray}
M_\Psi&=&M_G \quad (\mbox{noTP}), \\
M_\Psi &=& 1 \mbox{ TeV} \quad (\mbox{lightTP}).
\end{eqnarray}
In the first model, that we call not top partners (noTP), the decay
into top partners, either singly or in pairs, is not kinematically
allowed and therefore this model reproduces the main features of the
model that is currently used to interpret LHC searches. The second model,
called light top partners (lightTP), is the benchmark model for a
realistic CHM, in which top partners are expected to
be relatively light and therefore the decay into top partner pairs, or
in association with a SM top or bottom, is kinematically allowed. Our
choice of 1 TeV for the top-partner mass 
is a compromise between what one would expect from
the observed value of the Higgs mass and the current limits on top
partners. The other two top partners are almost degenerate
\begin{equation}
M_{T^\prime}\approx M_B \approx 1.13 \mbox{ TeV}.
\end{equation}
This model has all the features we discussed in the
introduction: a large gluon width, a small decay fraction into
$t\bar{t}$ and a large one into $t\bar{t} X$ -where $X$ stands for two SM
gauge or Higgs bosons-. In order to disentangle the different effects
we will also consider other three benchmark models in which we
artificially modify some of the couplings to highlight some of the
relevant features.
These models are denoted by  
\begin{center}
\begin{tabular}{ll}
lightTP$_{\mathrm{narrow}}$&
$M_\Psi=1$ TeV,
$\Gamma=\Gamma_{\mathrm{noTP}}$,
\\
lightTP$_{\mathrm{notop}}$&
$M_\Psi=1$ TeV, $g_{Gtt}=0$,
$\Gamma=\Gamma_{\mathrm{noTP}}$,
\\
singleTP$_{\mathrm{narrow}}$&
$M_\Psi=M_G/2$, $g_{Gtt}=g_{Gb_L b_L}=0$,
\end{tabular}
\end{center}
In all three cases we have rescaled the couplings of the composite
(and $t_L$, $b_L$) quarks to have a narrow resonance (explicitly we
have fixed the width to the one in the noTP model). Thus, the large
width effect is removed in these models. The
$\mathrm{lightTP_{narrow}}$ model is a narrow-resonance version of
lightTP, with quite similar decay patterns. In our second benchmark
model, $\mathrm{lightTP_{notop}}$, we have further set to zero the
couplings to the top (and re-scaled again the couplings to the top
partners to keep the same width) to ensure decay only to top
partners. Finally, in the $\mathrm{singleTP_{narrow}}$ model
we have chosen the top partner mass to
favour single top partner production. Since the decay into pairs of
top partners is kinematically forbidden, the width is relatively
small. However, the fact that the RH top is fully composite leaves a
small \textit{BR} into a top partner and a top or bottom quark. In order to
increase the \textit{BR} into these channels we set the couplings to the top
and left-handed bottom to zero and re-scale the couplings to
$tT$ and $bB$ to keep the original width.
As an illustration we provide in Table~\ref{benchmarks:table} 
the values of the decay branching fractions and the
width of the heavy gluon to the different fields for a reference mass
$M_G=2.5$ TeV.
\begin{table}[ht]
\begin{tabular}{lcccccc}
Model  & BR($G\to tt$) & BR($G\to bb$) & BR($G\to \Psi \Psi
$) &
BR($G\to \Psi \psi$) & $\Gamma_G/M_G$
\\
\hline
noTP  & 0.92 & 0.01 & 0 & 0 & 0.1
\\
lightTP & 0.15 & 0.004 & 0.75 & 0.08 & 0.65
\\
$\mathrm{lightTP_{narrow}}$ & 0.14 & 0.01 & 0.7 & 0.08 & 0.1
\\
$\mathrm{lightTP_{notop}}$ & 0 & 0.01 & 0.82 & 0.09 & 0.1
\\
$\mathrm{singleTP_{narrow}}$ & 0 & 0.007 & 0 & 0.94 & 0.14
\end{tabular}
\caption{Relevant heavy gluon parameters in our benchmark models for
  $M_G=2.5$ TeV. \label{benchmarks:table}} 
\end{table}

\section{LHC signatures} \label{sec:signatures}

\begin{figure}[htb]
\begin{center}
\begin{tabular}{cc}
\includegraphics[width=3.2in]{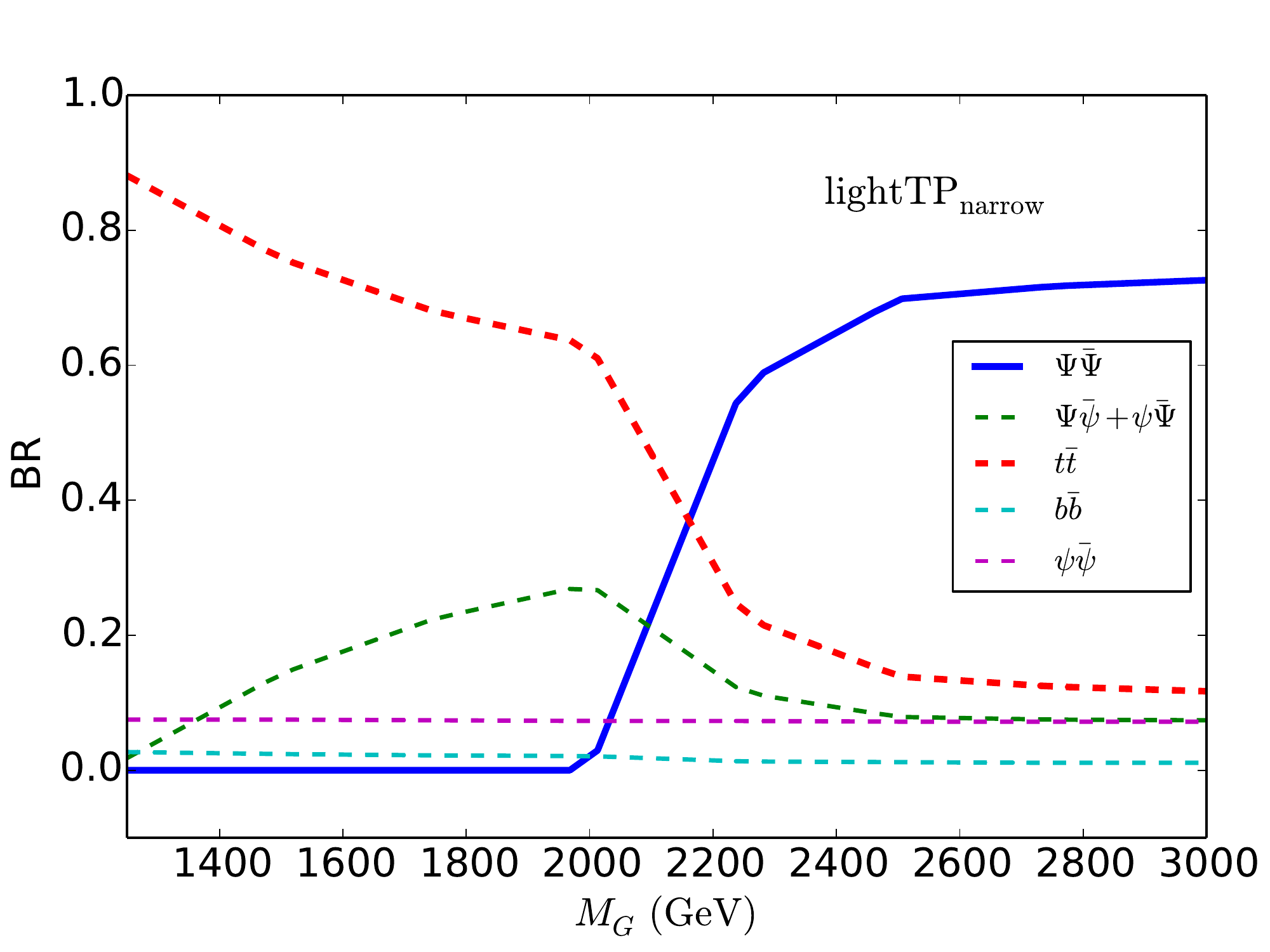}
\end{tabular}
\caption{The heavy gluon branching ratios in the lightTP model for all
  channels. The blue
solid line corresponds to the $\Psi\bar \Psi$ final state, with 
$\Psi = T^\prime, B, X_{2/3}^\prime, X_{5/3}$; the red dashed line to
the $t\bar t$; the green 
dashed line to the $T^\prime \bar t + t\bar T^\prime$ and $B\bar b + b\bar B$; the
cyan and purple dashed line correspond, respectively, to $b \bar b$
and light quarks.}
\label{fig:br}
\end{center}
\end{figure}

All the models analyzed in this work have one thing in common -- when
kinematically allowed, the heavy gluon has a large decay rate into
pairs of top partners or a top partner and a top or bottom quark. We
show as an example the branching ratios for heavy gluon decays in
Figure~\ref{fig:br} as a function of $M_G$ for the lightTP
model. Moreover, the total width of $G$ turns out to be generically in
the range of 50$\%$ to 80 $\%$ of the $G$ mass, which makes it a broad
resonance and extremely challenging to discover. As stressed above,
this is in sharp contrast with previous studies of heavy gluon searches
at the LHC in which $G$ was assumed to be not too broad and to
decay predominantly to $t\bar t$.  

Here, we are going to show the main phenomenological differences
between the heavy gluon in the \mch scenario and the heavy gluon in
the RS KK gluon scenario in regard to final states from their decays
and their kinematical reconstruction. We base our discussion on the
class of models described in Section~\ref{sec:model}. 
 In order to disentangle the main effects that can make the heavy
 gluon elusive, in this Section we focus on the results at partonic
 level and postpone a discussion on the full reconstruction of the
 heavy gluon resonance until Section~\ref{sec:results}.  This will
 give us a good handle on how well the LHC $t\bar t$ resonance
 searches, designed for RS KK gluon model in mind, will do at covering
 the much more general \mch parameter space.  The parton-level
 analysis also gives us some useful information that is less sensitive
 to a particular experiment or reconstruction method.

\begin{figure}[htb]
\begin{center}
\begin{tabular}{cc}
\includegraphics[width=3.2in]{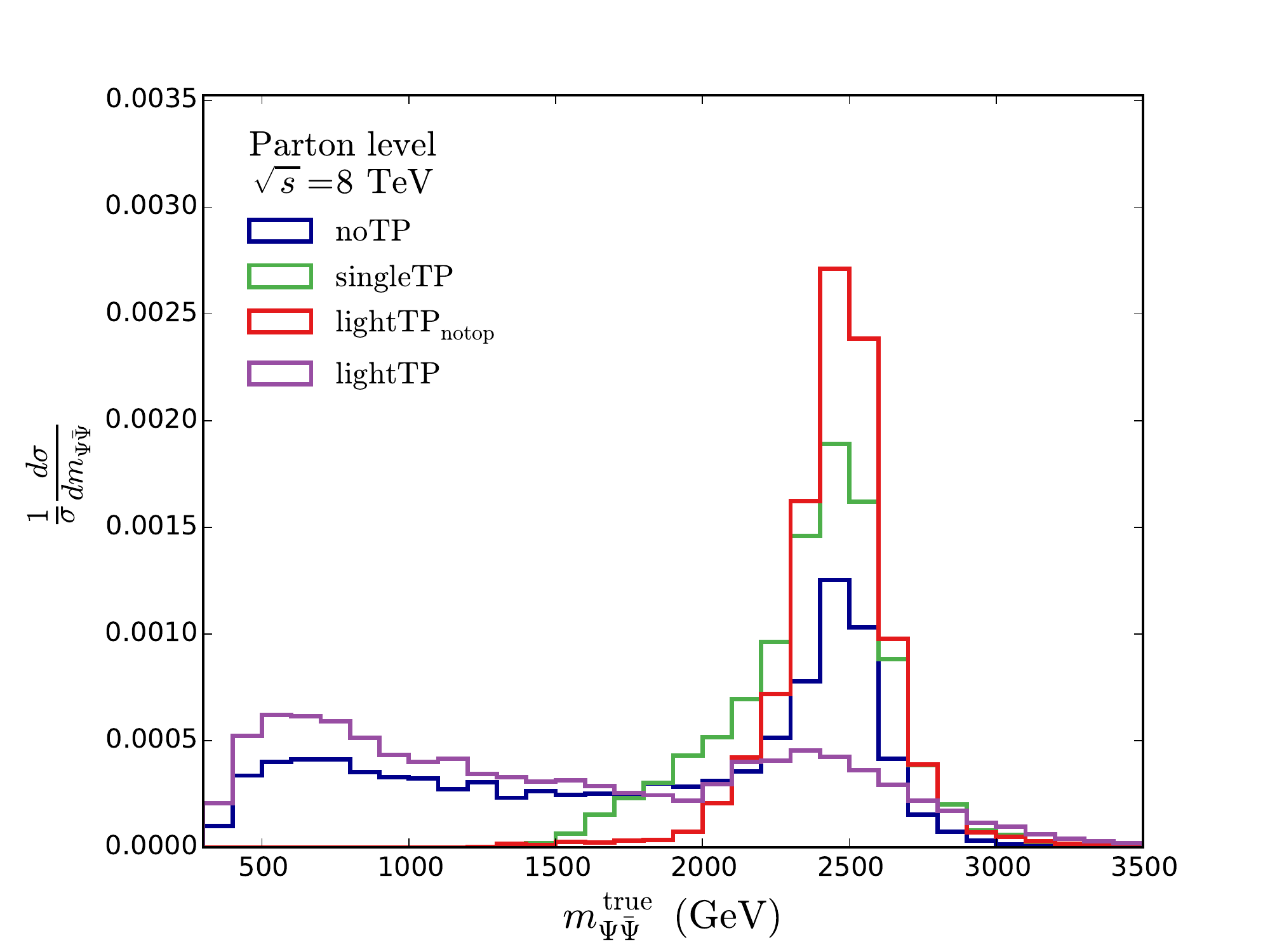} &
\includegraphics[width=3.2in]{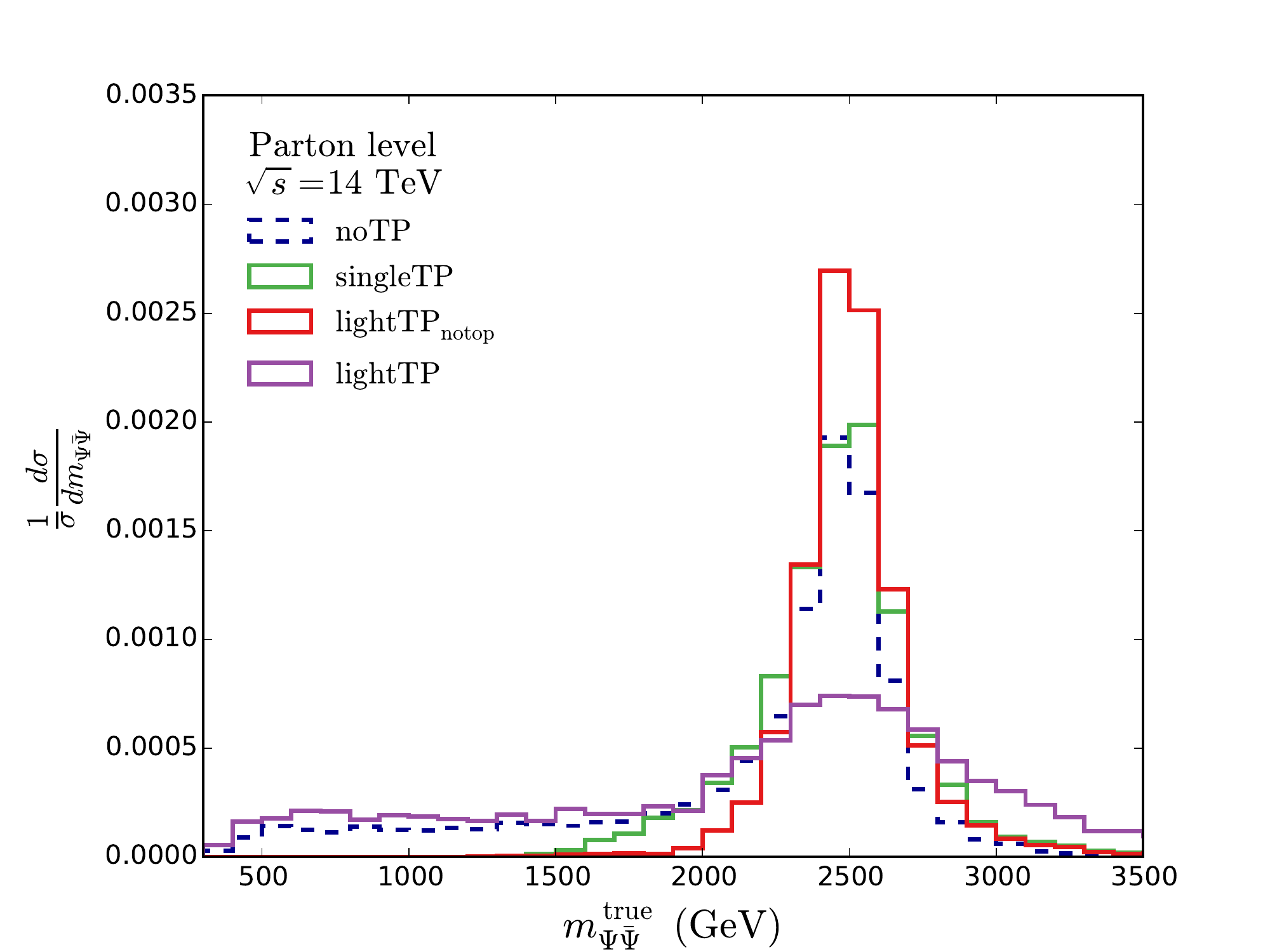} \\
\end{tabular}
\caption{The true invariant mass distribution of the decay products of
  $G$  in $pp\to G \to \psi\bar \psi, \Psi\bar \psi, \Psi\bar \Psi$
  for $M_G=2.5 \TeV$ with $\sqrt{s}=8\TeV$ 
(left) and $\sqrt{s}=14\TeV$ (right). 
} 
\label{fig:mtt_twoprong_decays}
\end{center}
\end{figure}

 Starting from the Lagrangian, described in detail in the Appendix
 (see Eq.~(\ref{Lag:MCH45})), one can derive the phenomenology of the
 heavy gluon. Since the coupling of the heavy gluon to ordinary gluons
 vanishes at tree level, the main rate of production at the LHC comes
 from the Drell-Yan process $q\bar q \to G $. The differential cross
 section as a function of $m_{\Psi\bar \Psi}^{\rm true}$ at LO is
 presented in Figure~\ref{fig:mtt_twoprong_decays} for $M_G=2.5\TeV$
 produced at the LHC with $\sqrt{s}= 8 \TeV$ (left panel) and
 $\sqrt{s}=14 \TeV$ (right panel). The invariant mass for the heavy
 gluon in the noTP model is shown for comparison in the same figure.
 The invariant mass is calculated from the decay products of $G$ using
 truth level particles directly from Monte Carlo. The shape of the
 distribution for the noTP model simply corresponds to the
 Breit-Wigner factor convoluted with the corresponding parton
 distribution functions (PDF). By
 contrast, the invariant heavy gluon mass for the lightTP model is in
 average quite low and spread out. This is a direct consequence of the
 large width of the heavy gluon together with enhancement of PDF's at
 low x, which lead to significant departure from the narrow-width
 approximation. In this case, a search for the heavy gluon would be
 quite a challenge, if not impossible.  The invariant mass
 distributions of the narrow models, $\rm lightTP_{narrow}$ and $\rm
 singleTP_{narrow}$,  are more sharply peaked around the heavy gluon
 mass and the distributions are more symmetric.   The relatively small
 width of the heavy gluon for the narrow-resonance models suggests a
 new promising strategy for discovering the $G$, by searching for
 resonances in the invariant mass distribution of top partner pairs. 

Considering the pattern of top partner decays, we can identify the
following final channels: 
\begin{eqnarray}
  p p \to &G& \to X_{5/3}\bar X_{5/3} \to (t W^+) (\bar t W^-) \label{Eq:decay_topologies_1} \nonumber \\
    p p \to &G& \to X_{2/3}\bar X_{2/3} \to ( t Z + t h ) (\bar t Z + \bar t h)\nonumber \\
   p p \to &G& \to T\bar T \to ( t Z + t h+ b W^+ ) (\bar t Z + \bar t h+ \bar b W^-)\nonumber \\
   p p \to &G& \to B\bar B \to (t W^-) (\bar t W^+)  \label{Eq:decay_topologies_2}\\
   p p \to &G& \to t \bar T + b\bar B +c.c.  \to  t(\bar t Z + \bar t h+ \bar b W^- ) +  b (\bar t W^+)+c.c.\nonumber \\
   p p \to &G& \to (t\bar t +b\bar b + q\bar q)\nonumber
\end{eqnarray}
As can be seen here, most of these processes lead to events with top
quark pairs and $W$, $Z$ or Higgs bosons in the final state.  This is
interesting as in the searches for RS KK gluon resonances,  the heavy
gluon mass is kinematically reconstructed only from the tagged top
quarks and the extra Higgs or vector bosons are not identified.
Assuming that the top pairs can be reconstructed with good quality,
one might expect the additional heavy bosons to have great qualitative
impact on the reconstruction of the heavy gluon resonance at the
LHC. Indeed that is case in some special kinematical regions, as we
shall see.

\begin{figure}[htbp]
\begin{center}
\begin{tabular}{cc}
\includegraphics[width=3.2in]{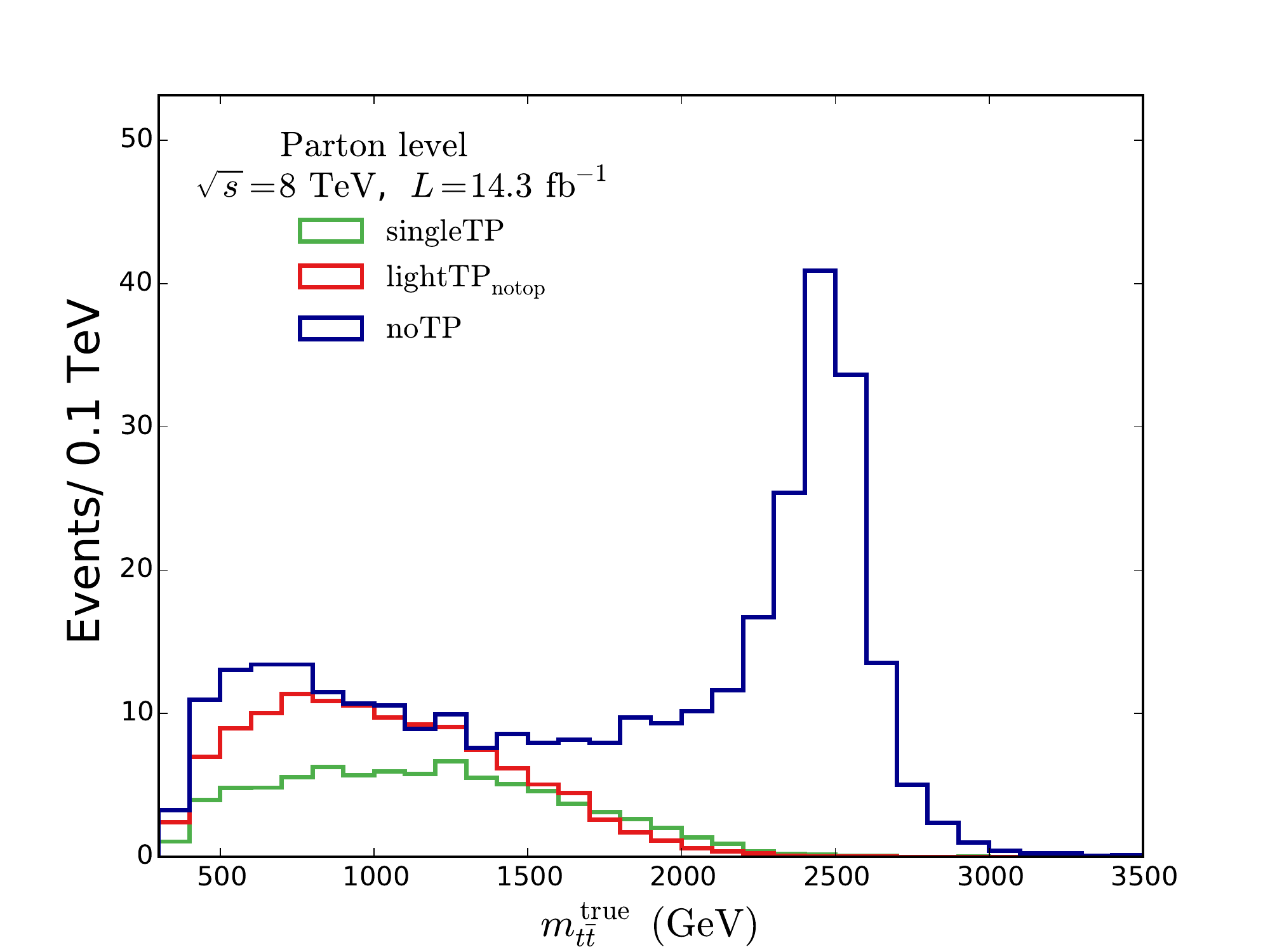} &
\includegraphics[width=3.2in]{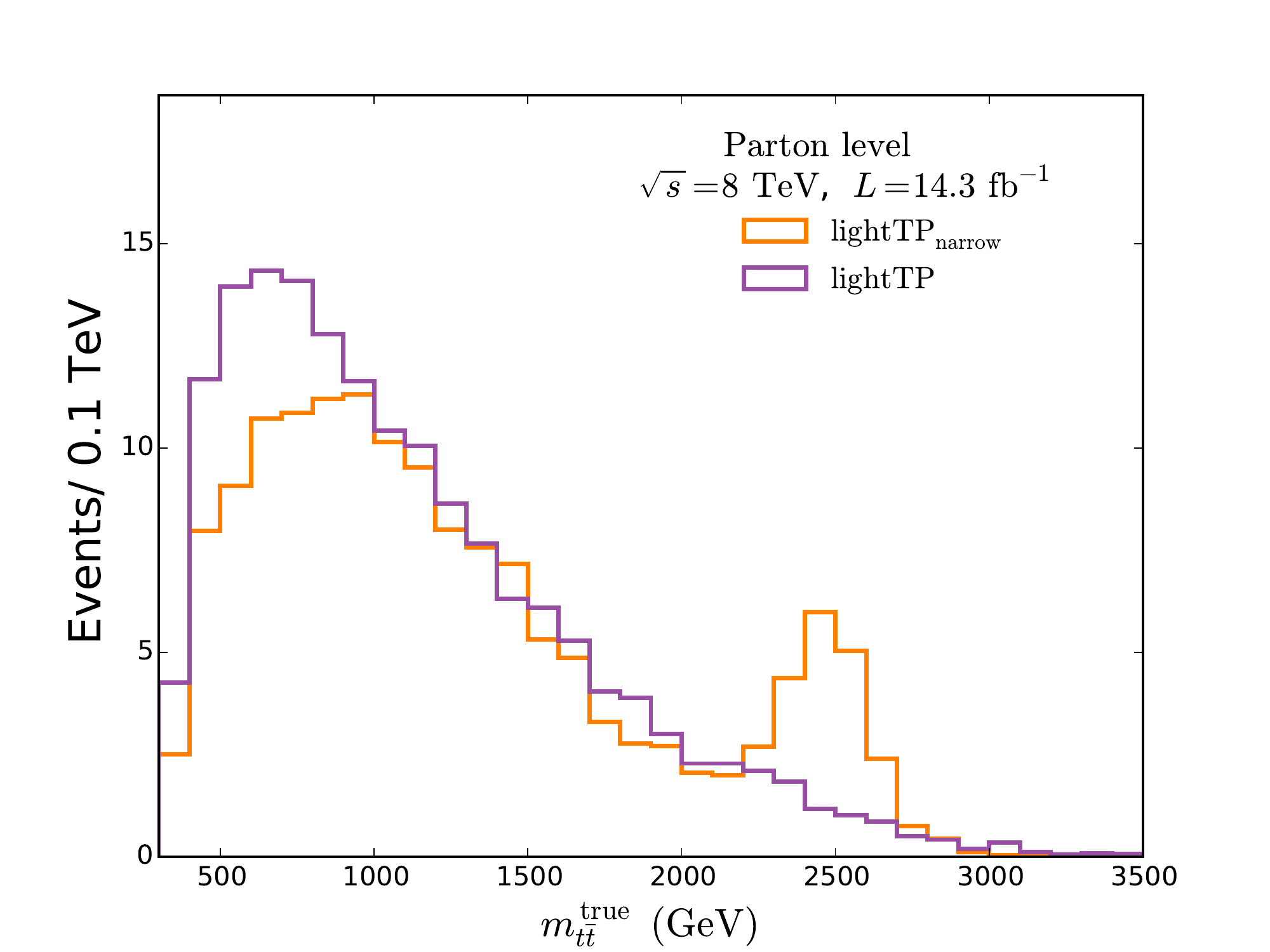}
\end{tabular}
\caption{True (partonic) invariant mass for the $t\bar t$ pairs in $
  p\bar p \to G \to t\bar t + X$ for the noTP, $\rm lightTP_{notop}$
  and $\rm singleTP_{narrow}$ models (left panel) and $\rm
  lightTP_{narrow}$ and $\rm lightTP$ models (right panel). The plots
  correspond to  $M_G= 2.5 \TeV$ at the 8 TeV LHC. 
}\label{fig:mtt_part_8tev}
\end{center}
\end{figure}

\begin{figure}[htbp]
\begin{center}
\begin{tabular}{cc}

\includegraphics[width=3.2in]{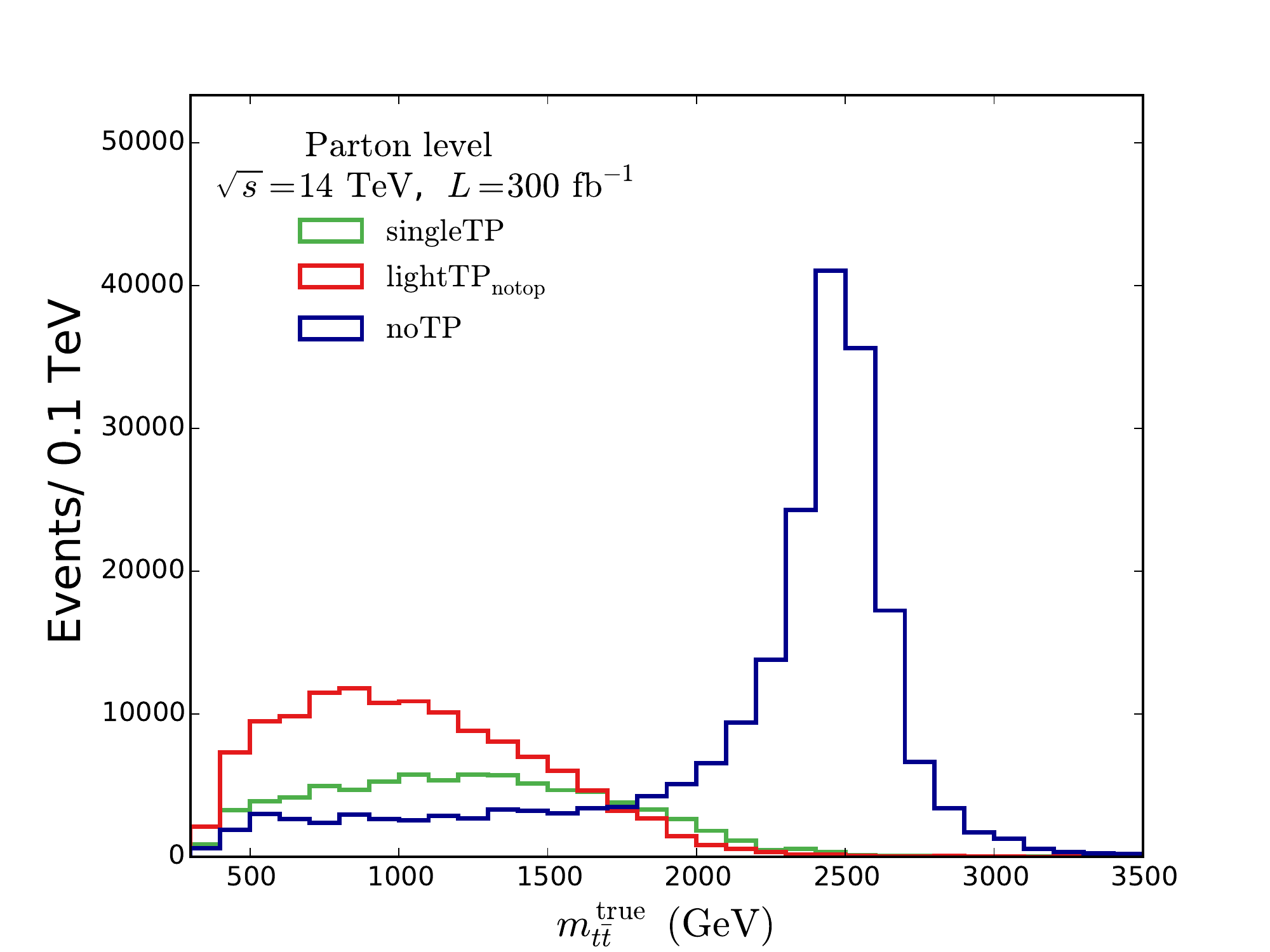}  &
\includegraphics[width=3.2in]{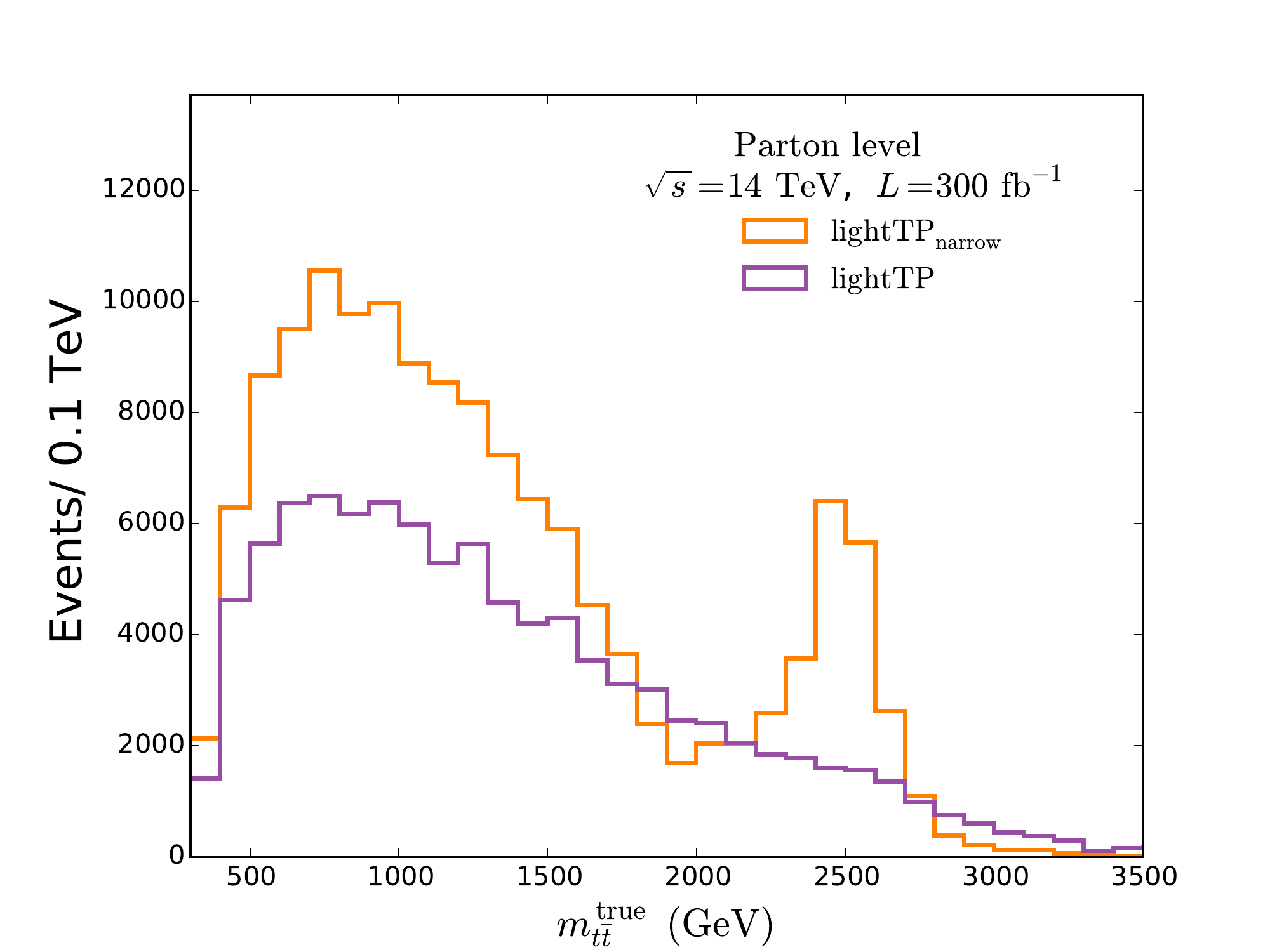}
\end{tabular}
\caption{True (partonic) invariant mass for the $t\bar t$ pairs in $
  p\bar p \to G \to t\bar t + X$ for the noTP, $\rm lightTP_{notop}$
  and $\rm singleTP_{narrow}$ models (left panel) and $\rm
  lightTP_{narrow}$ and $\rm lightTP$ models (right panel). The plots
  correspond to  $M_G= 2.5$ TeV at the 14 TeV LHC. 
}
\label{fig:mtt_part_14tev}
\end{center}
\end{figure}

We show in Figure~\ref{fig:mtt_part_8tev} the $m_{t\bar t}$ distribution
 at truth level at the 8 TeV LHC for the extreme benchmarks
 (noTP, lightTP$_\mathrm{notop}$ and singleTP$_\mathrm{narrow}$) in
 the left panel and for the  
 realistic benchmarks (lightTP$_\mathrm{narrow}$ and lightTP) in the right one. 
The plots have been normalized to an
 integrated luminosity of $14.3 ~\rm fb^{-1}$. The corresponding results
 for the LHC14 and $300~ \rm fb^{-1}$ of data are shown in
 Fig.~\ref{fig:mtt_part_14tev}. 
 We see that the shape of simple Breit-Wigner type resonance is
 largely distorted especially for all models with new channels. These
 huge effects can be easily understood by the peculiar topology of top
 partner decays, which, as shown in Eq.~(\ref{Eq:decay_topologies_1}),
 leads to extra particles in the final state besides the top pairs.
 The invariant mass of most of the top-quark pairs produced from the
 $G$ decays is, therefore, much smaller than the $G$ mass as can be
 clearly seen in Figures~\ref{fig:mtt_part_8tev} and~\ref{fig:mtt_part_14tev}. 
 These results suggest that a purely $t\bar t$ resonance search  would
 not be efficient at reconstructing the heavy gluon in the more
 general case of the \mch parameter space. We investigate this effect
 in the reconstructed sample in Section~\ref{sec:results}. 

\section{Current bounds} \label{sec:exp_limits}
Both the ATLAS and CMS collaborations search for heavy resonances decaying in $t\bar t$ in both 7 and 8 TeV data
using the reconstructed  top quark pair invariant mass ($m_{t\bar
  t}$). These searches are sensitive to new resonances decaying to top
quark pairs of various resonance widths, including narrow $Z'$ bosons
and broader heavy gluons. In the case of high-mass resonances, the use of
jet substructure techniques to efficiently capture boosted top quarks
or W bosons is crucial to achieve a good background rejection and
efficient event reconstruction.  All possible decay channels are
considered in these studies: all-hadronic, lepton + jets and dilepton
topologies. 

Searches in the dilepton decay mode are performed by both
ATLAS~\cite{Petteni:2011np} and CMS~\cite{CMS:2014rna}. In the former
case the results are based on the
LHC7 data, and the exclusion limits reach to roughly 1 TeV, depending
on the signal model considered. In the latter case 8 TeV data is used
and the limit goes up to $1.8$ TeV. Searches in the all-hadronic decay
mode make use of top-tagging techniques. The latest results from CMS
include $19.6 \,{\rm fb}^{-1}$ of LHC8 data and RS KK gluons are
excluded up to 1.8 TeV~\cite{CMS:2013vca}. The latest results from
ATLAS analyze $4.7 \,{\rm fb}^{-1}$ of LHC7 data, excluding KK gluon
with masses below 1.62 TeV~\cite{Aad:2012raa}. 

 Searches in the lepton + jets channel are also carried out by
 ATLAS~\cite{TheATLAScollaboration:2013kha} and
 CMS~\cite{Chatrchyan:2013lca} with LHC8 data. 
ATLAS excludes KK gluons with masses
 up to 2.0 TeV using $14.3 \,{\rm fb}^{-1}$ of integrated
 luminosity. CMS, using $19.6 \,{\rm fb}^{-1}$, excludes KK gluons up
 to 2.5 TeV, with a slight difference in the signal model with respect
 to ATLAS. To date these searches yield the strongest bounds on the RS KK
 gluon.

 Since our focus is on disentangling the different effects of top
 partners on the heavy gluon searches, and not on improving the
 existing analysis, in this work we will consider only one of these
 searches, namely the lepton+jets analysis performed by the ATLAS
 collaboration. Because ATLAS and CMS employ somewhat different top
 reconstruction methods and have slightly different signal
 models\footnote{The main source of the quantitative difference is
   that the ATLAS study only considers production of KK gluon
   resonance assuming a LO KK gluon production cross section, whereas
   CMS applies a flat K-factor of 1.3 to account for higher order
   effects. Meanwhile, the CMS observed limits on the KK gluon mass
   turned out to be higher than their expectation, whereas ATLAS
   observed limits were somewhat below their expectation.}, the
 results based on either of these methods would differ quantitatively,
 but qualitatively the shape of the distributions are expected to be
 similar.

We would like to make an important comment regarding the interplay
between vector resonance and top partner searches.  
The standard production of top partners at the LHC is either pair
production through QCD interactions or single production involving
model-dependent couplings.
Several experimental searches for top partners have been
performed by ATLAS~\cite{Aad:2014efa} 
and CMS~\cite{Chatrchyan:2013uxa,Chatrchyan:2013wfa} with LHC7 and LHC8 data. 
The resulting lower bounds on the top partner masses depend on the
channel and the assumptions on the branching ratios but they are typically in the
600-800 GeV region. These analyses are quite inclusive in the final
states but mostly consider QCD pair production as they are not yet
sensitive to single production
(see~\cite{Aguilar-Saavedra:2013qpa,Matsedonskyi:2014mna} for a
discussion of the 
relevance of single production in these searches). 
In the presence of a heavy gluon, there is an extra contribution to pair
and single production of top partners mediated by the s-channel
exchange of $G$. Thus, the current limits on top partners are expected
to be more stringent than the ones reported above.
For the benchmark models considered here, however,
the event rate from a heavy gluon resonance contributes to the signal
without running into serious conflict with existing experimental
bounds. Choosing a mass of 1 TeV for the lightest top partners yields
an estimate of 20.7 fb for the total cross section (conservatively
including a K-factor of 1.3) while the current limit from CMS at this mass
point is 23.7 fb.

\section{Method} \label{sec:analysis}

We will now describe the analysis procedure that we have followed
in generating and analyzing the signal events for heavy gluon
production at the LHC8 and LHC14. Throughout our analysis, we
adhere to the search strategies developed by
ATLAS~\cite{TheATLAScollaboration:2013kha}, designed with the RS KK
gluon model in mind, as closely as possible. 
The justification for the choices of specific analysis cuts as well as
the use of the statistical tools for exclusion limits employed here
are the same as those employed by ATLAS. In particular, we focus on
the lepton+jets reconstruction mode. 

\subsection{Event generation} \label{sec:event_generation}

The implementation of the signal models was performed using
\texttt{Feynrules}~\cite{Alloul:2013bka}. 
The production of samples was done with
\texttt{MadGraph/MadEvent}~\cite{Maltoni:2002qb}~v.1.5.3  interfaced with
\texttt{Pythia}~\cite{Sjostrand:2006za}~v.~6.426 for showering and
hadronization.  For the SM $t\bar t$ production, we use MLM
matching~\cite{Mangano:2006rw} with up to one additional hard parton.
We use the default tunes of Pythia for the hadronization and
underlying event model parameters, 
while for the matching scale in the $t\bar t$ sample we use $Q_{\rm
  cut} = 30 \GeV$ in both cases.\footnote{For the parton separation
  parameter of the MLM matching procedure in \texttt{MadGraph}, 
we use $ \texttt{xqcut}= 20 \GeV$.}
Our MadGraph samples assume the {\sc CTEQ6L1}~\cite{Nadolsky:2008zw}
PDF sets.  While the signal cross sections
are computed at LO, the background cross section is obtained with
\texttt{MadGraph} normalized to the theoretical NNLO cross section of
Ref.~\cite{Czakon:2013goa}. We perform jet clustering using the {\sc
  Fastjet}~\cite{Cacciari:2011ma} 
implementation of the anti-$k_T$ algorithm~\cite{Cacciari:2008gp}.  

The main background process is the irreducible SM $t\bar t$  after
imposing all cuts described in 
the next Section. The $W$+jets  and pure multi-jet QCD background
events are difficult to simulate reliably, but we expect that they are
efficiently suppressed by our cut procedure, in particular by the
mini-Isolation cut and top tagging.

Not having at our disposal a reliable tool to estimate the response of
the detector, a semi-realistic simulation of the hadronic final states
would not be useful. Therefore we decided not to include the  effects
of pileup and underlying event in our analysis, as well as detector
effects, and we stopped at the hadronization level. We treat electrons
and muons together in order to get an estimate of the kinematical
acceptance. Neither do we attempt to include uncertainties of the 
background. Consequently, this article does not present a fully
realistic analysis but rather demonstrates the impact of new
vector-like quarks on heavy gluon searches. 

 \subsection{Analysis details} \label{sec:event_selection}

The analysis of the samples of simulated events is performed on the
 stable final-state particles using a custom analysis tool aimed at
 mimicking the ATLAS $\ell$+jets search. The event selection is
 designed to tag $p p \to t\bar t$ events with subsequent decay $t\bar
 t \to b\bar b j j \ell \nu_\ell$. The expected kinematics of the
 top-quark decay products is characterized by two event topologies. In
 the first category, the $t\bar t$ pair is produced near the kinematic
 threshold, resulting in a topology where each parton is matched to a
 single jet (resolved topology). In the second category, each top
 quark is produced with a high Lorentz boost, resulting in collimated
 decay products that may be clustered into a single jet (boosted
 topology). The transition between the resolved and boosted
 topologies occurs around $m_{t\bar t} = 1 \TeV$. The resolved and
 boosted selections are now discussed in detail. 

The physics object selection criteria closely follow those used in the ATLAS $\ell$+jets  search, the main exceptions being the treatment of detector effects.
Charged leptons are required to be \textit{mini-isolated}~\cite{Aad:2013nca},
\ba
{\miso} \equiv \frac{p_T^\ell}{p_T^{\rm cone}} > 0.9\, ,\,\,\,\,\,\,  \Delta R (\ell, \,{\rm track}) < \frac{10{\rm \, GeV}}{p_T^{\ell}}, \label{eq:lepSel}
\ea
where mini-ISO is the lepton isolation observable of Ref.~\cite{Kaplan:2008ie} and $p_T^{\rm cone} $ is scalar sum of all the charged tracks with $p_T>1\,$GeV, including the hard lepton, that fulfill the $\Delta R (\ell, \, \rm track)$ requirement shown in Eq.~(\ref{eq:lepSel}).

Small-radius jets are clustered from all final-state particles except
neutrinos using the anti-$k_T$ algorithm with $R=0.4$. Only jets with
$p_T > 25 \GeV$ and $|\eta|<2.5$ are used. Large-radius jets are
clustered in a similar way, but with a large radius $R=1.0$. These
large-radius jets are required to have $p_T > 350 \GeV$ and $|\eta|<2.5.$

We use a simple-minded algorithm for $b$-tagging in which an
anti-$k_T$ $R=0.4$ jet is matched to the corresponding $b$ parton. The
typical performance of experimental $b$-tagging was simulated by
applying  a flat $b$-tagging efficiency of 0.7.  For the purpose of
this analysis, we define a transverse missing energy vector,
$E_T^{miss}$, to be the vector sum of all the neutrino transverse
momenta in the event. The transverse mass is defined as $m_T = \sqrt{2
  p_T E_T^{miss}(1-\cos \Delta \phi)}$, where $p_T$ is the transverse
momentum of the charged lepton and $\Delta \phi$ is the angle in the
transverse plane between the charged lepton $p_T$ and the missing
transverse momentum.  

Events are preselected by requiring exactly one mini-isolated electron or muon with  $p_T > 25 \GeV$, $E_T^{miss} > 20 \GeV$ and $E_T^{miss}+ m_T > 60 \GeV$.
Events are also required to pass either the boosted or resolved selections. In the \textit{Boosted Selection}, events contain at least one anti-$k_T$ $R = 0.4$ jet and at least one anti-$k_T$  $R = 1.0$ jet. The highest $p_T$ small radius jet within a distance $\Delta R_{j\ell} < 1.5$ from the lepton is deemed the $b$-jet of the leptonically decaying top. The fat jet must be well separated from the lepton and selected $b$-jet: $\Delta \phi_{J\ell} > 2.3$ and $\Delta R_{J j}>1.5$.  Two additional requirements on the substructure of the fat jet are made, the so-called \ADTT~\cite{Aad:2013nca}, which consists of the following cuts:
\be
        \sqrt{d_{12}} > 40.0 \GeV , \,\,\,\,\, m_j > 100 \GeV \label{eq:d12}\,.
\ee
The $m_j$ is the fat jet mass~\footnote{Strictly speaking, ATLAS uses $m_j^{\rm trim}$, which is the trimmed fat jet mass with the trimming parameters $R_{\rm trim} = 0.3$ and $f = 0.05$
(see Ref.~\cite{Krohn:2009th} for more details).}, and $\sqrt{d_{12}}  = {\rm min}(p_{T,1}, p_{T,2}) \times \Delta R_{12}$ is the $k_T$ measure at the last step of large-radius jet clustering with a  $k_T$ algorithm, where  $p_{T,i}$  are the transverse momenta of the two subjets at the last step of fat jet  clustering and $\Delta R_{12}$ is the plane distance between them. Boosted top quark decays are characterized by symmetric splittings $\sqrt{d_{12}} \approx m_{t}/2$, whereas background QCD jets tend to have much smaller $d_{12}$.

In the \textit{Resolved Selection}, events contain at least four small-radius jets with $|\eta|<2.5$ and $p_T > 25 \GeV$, or only three small-radius jets if one of those jets has a mass greater than $60 \GeV$.  The cuts and other kinematical constraints are summarized in Table~\ref{tab:sel_cuts}.

{\small
\begin{table}[htbp]
\begin{center}\hspace*{.1cm}
\begin{tabular}{|l||c|c|}
\hline
   \textbf{Selection}  &  \textbf{Cuts}      \\
\hline
\hline
                    &   lepton                   $p_T > 25 \,{\rm GeV},\,|\eta|<2.5$ \\
  Kinematic    &  $\ge$ 2 jets              $p_T > 25 \,{\rm GeV},\,|\eta|<2.5$ \\
    and     &  tagged $b$-jets         $\ge$ 1\\
  acceptance                    &  missing energy ($\nu$)    $E_T^{miss} > $ 20 GeV        \\
        &transverse mass $m_T + E_{\rm T}^{\rm miss} > 60 \GeV$\\
                    &  lepton isolation   mini-ISO$> 0.9$         \\

\hline \hline
 Resolved  selection   &  $\ge$ 4 jets     $p_T > 25 \,{\rm GeV},\,|\eta|<2.5$  or \\
 for \#jets$>2$,  &  3 jets              $p_T > 25 \,{\rm GeV},\,|\eta|<2.5$ \& \\
 1 $b$-jet required & $\ge 1$ jets $m_j > 60 \GeV$\\

  \hline
    Boosted   selection     &  $m_{\rm jet} >100 \GeV$ \& $ \sqrt{d_{12} }> 40 \GeV$    \\
    1 $b$-jet+ 1 $R=1.0$-jet&        $p_T^{\rm jet}>$ 350 GeV   \,\& \, $|\eta_{\rm jet}| < 2.5$  \\
    &  $\Delta \phi({\rm jet},l) > 2.3$  \,\& \, $\Delta R ({\rm jet}, b) >1.5$\\
 \hline
\end{tabular}
\vspace*{.2cm}
\caption{Selection cuts in the semileptonic $t\bar{t}$ channel.
}
\label{tab:sel_cuts}
\end{center}
\end{table}
}

The $t\bar t$ invariant mass, $m_{t\bar t}$, is computed from the
four-momenta of the two reconstructed top quarks. For the
leptonically decaying top quark, the longitudinal momentum of the
neutrino, $p_z$,  is computed by imposing the $W$ boson mass
constraint, ($M_{l \nu} = M_W = 81 \GeV $), and solving the resulting
quadratic equation. This information is sufficient to reconstruct the
neutrino momentum, modulo a quadratic ambiguity.  In the case that the
solutions are complex, the magnitude of $E_T^{miss}$ is reduced to the point
where $m_T (l, E_T^{miss} ) = m_W $. In the case where we obtain two
solutions, both solutions are tried.

For the resolved reconstruction, a $\chi^2$ algorithm is used to
determine the correct assignment of jets to top quark candidates,
using as constraints the top quark and $W$ boson masses and other
kinematic properties of the signal process. All possible permutations
for three or more small radius jets are tried and the permutation with
the lowest $\chi^2$ is used to compute the $m_{t\bar t}$
distribution. This method is optimized for events containing tops and
$W$'s which are not too energetic, as in the case of the $t\bar t$
events with top quark invariant masses smaller than about 1 TeV. In
this case, leptons will be isolated, and there is good match between
jets and parton momenta. 

For the boosted reconstruction, there is no ambiguity in the
assignment of jets. The hadronically-decaying top quark is taken from
the fat jet, while the leptonically decaying top quark momentum
is formed from the neutrino solution, the lepton and the selected
small-radius jet. The $m_{t\bar t}$ distribution is used for signal
discrimination, after combining the resolved and boosted analyses.

\subsection{Statistical procedure} \label{sec:statistical_procedure}
To determine the expected reach for the five benchmark models in
Section~\ref{sec:model}, we use a binned Bayesian approach with a flat,
positive prior on the signal cross section.   We assume that the
probability of measuring $n$ events in statistically uncorrelated bins
is given by a Poisson distribution 
\begin{equation}
P(\{n\}|S,B) =  \sum_{\rm bins} \left[\frac{(S_i+B_i)^{n_i}}{n_i!}
  e^{-(S_i+B_i)} \right]\,, 
\end{equation}
where $B_i$ and $S_i \equiv  \sigma_{\rm sig}\epsilon_{i} \mathcal{L}$ 
are the number of expected background and signal events in each
bin. Here we regard $\sigma_{\rm sig}$ as a free parameter in order to
consider different signal production rates and fix $B$ and
$\epsilon_{i}$ according to our expectations based on the Monte Carlo
distributions. 
 An upper limit for $\sigma_{\rm sig}$  at confidence level ${\rm
   CL}=1-\alpha$ can be constructed by integrating the posterior
 probability, 
\begin{equation} \label{eq:bayes}
{\rm CL} = 1- \alpha =\frac{ \int_{0}^{\sigma^{ \rm CL}} P(n |
  \sigma_{\rm sig} \epsilon_{\rm sig} \mathcal{L} ,B) d{\sigma_{\rm
      sig}}}{ \int_{0}^\infty P(n|\sigma_{\rm sig} \epsilon_{\rm sig}
  \mathcal{L},B) d\sigma_{\rm sig} }\,. 
\end{equation}
To obtain the expected limit on the signal cross section,  we solve
Eq.~(\ref{eq:bayes}) for $\sigma^{ \rm CL}$ assuming $n=B$ and $\alpha
=0.05$ ($95\%$ exclusion).

\subsection{Comparison with ATLAS benchmark analysis} \label{sec:atlas_benchm}

\begin{figure}[htb]
\begin{center}
\begin{tabular}{ccc}
\includegraphics[width=3.2in]{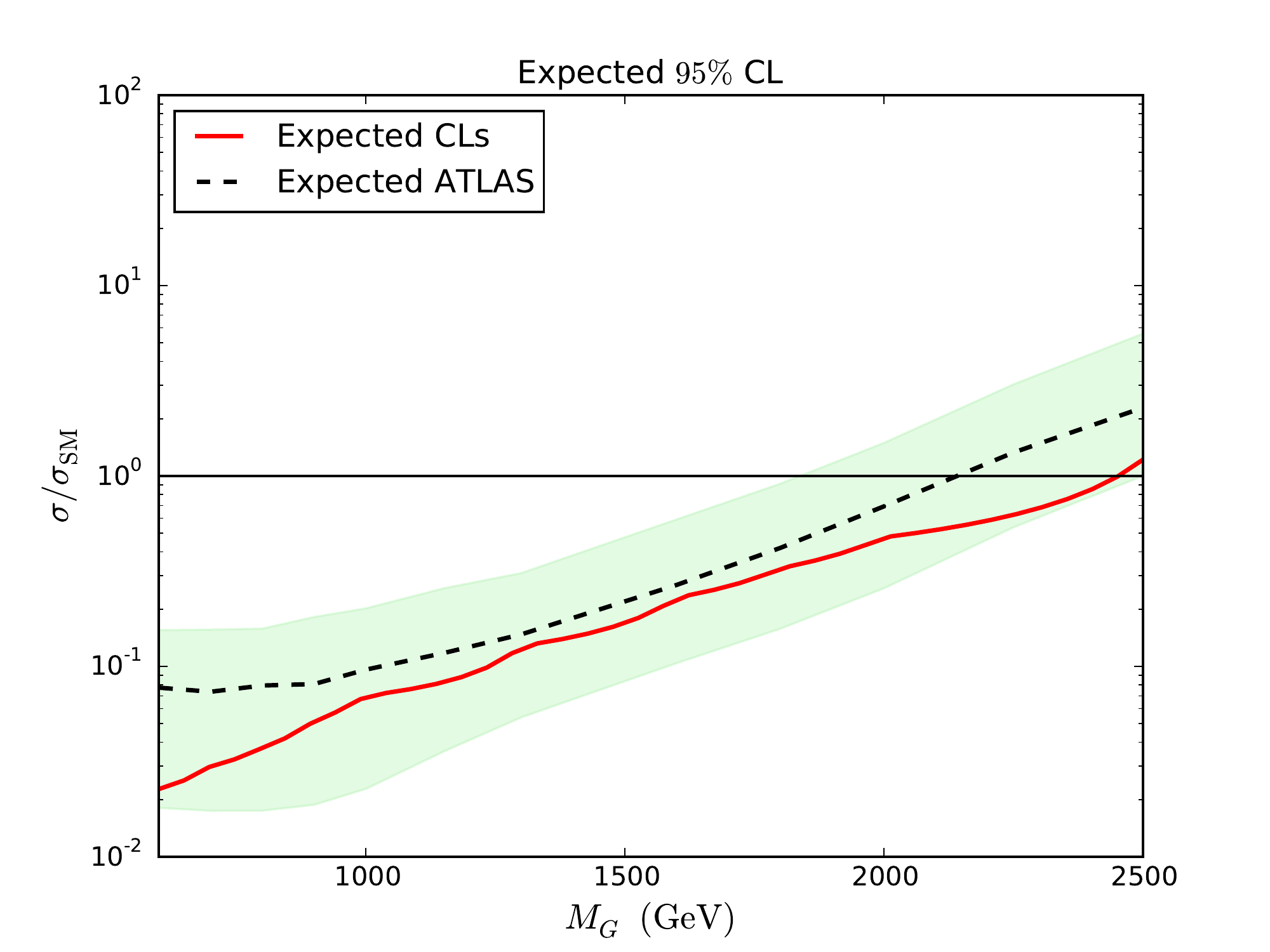} &

\end{tabular}
\caption{The $95 \%$ confidence level limit on the heavy gluon production cross section divided by the expected  $pp \to g_{KK}\to t\bar t$ production in the RS KK gluon model.}
\label{fig:cls_kk_atlas}
\end{center}
\end{figure}

ATLAS has searched for RS KK gluons using $14.3 ~{\rm fb}^{-1}$ of
 LHC8 data. As a sanity check, it is imperative for us to compare the
 results of our analysis with these published results for the same
 benchmark model before proceeding to apply our analysis to the \mch
 benchmark set. The comparisons shown in Figure~\ref{fig:cls_kk_atlas}
 and Table~\ref{tab:kkg_eff} indicate that we are indeed able to
 reproduce to a reasonable degree the results obtained by ATLAS for
 this benchmark model for the $\ell$+ jets analysis. 
 \begin{table}[h]
 \begin{center}
\begin{tabular}{ccc}\hline
\multicolumn{1}{l}{$g_{KK}$ mass [TeV]} & \multicolumn{1}{l}{Resolved selection} & \multicolumn{1}{l}{Boosted selection} \\\hline\hline
500  &   0.027  (0.0351$\pm$0.0029) &   0.0004 (0.00042$\pm$0.0001) \\
600  &   0.033  (0.0400$\pm$ 0.0032) &   0.0009  (0.00122$\pm$0.0003) \\
700  &   0.037 (0.0440$\pm$ 0.0032)&   0.004 (0.0039$\pm$ 0.0011)  \\
900  &   0.036  (0.0400$\pm$ 0.0032) &   0.019   (0.0170$\pm$ 0.0022)\\
1000  &   0.035  (0.0370$\pm$0.0028) &   0.026  (0.0242$\pm$0.0022)\\
1300  &   0.031  (0.0344$\pm$0.0024) &   0.039  (0.035$\pm$0.0021) \\
1600  &   0.032  (0.0304$\pm$0.0018) &   0.045  (0.039$\pm$ 0.004) \\
1800  &   0.030  (0.0289$\pm$ 0.0017) &   0.046  (0.042$\pm$ 0.005) \\
2000  &   0.030 (0.0286$\pm$ 0.0017)  &   0.050  (0.041$\pm$ 0.007) \\
2500  &   0.029  (0.0293$\pm$0.0017) &   0.045  (0.038$\pm$ 0.008)
\end{tabular}
\caption{Acceptance $\times$ efficiency for $G\to t\bar t$ samples in the $\mu$+jets channel. The ATLAS result is in parenthesis.}
\label{tab:kkg_eff}
\end{center}
\end{table}

Here it is important to remind the reader that our event generation
and analysis differs slightly from the procedure used by ATLAS. In
particular, our simulation of SM $t\bar t$ background includes only
the real emissions through matching with no contributions from the
virtual part of the NLO diagrams. These higher order effects could
have a great impact on the shape of the distributions, especially
around the region of high $t\bar t$ invariant masses.  As noted
before, 
we do not apply a full detector simulation. Furthermore, in our
statistical procedure we have neglected the effect of systematic
uncertainties in the background.  Therefore, some degree of
discrepancy can be expected. Notice, however, the small differences
will not affect the conclusion of this work, namely that semi-leptonic
$t\bar t$ searches rapidly loose sensitivity once the decay into top
partners are open. This is a generic statement in the sense that once
the signal becomes elusive the signal-to-background ratio reduces by
about two orders of magnitude, and this ratio is not expected to
change significantly after detector smearing or inclusion of
systematic effects. 

\section{Results}
\label{sec:results}
Now that we have introduced the main points of our analysis, we turn
to a quantitative discussion of the results for our MCH4$_5$
set-up. The extended scenario under consideration has fermionic
partners of the top together with the heavy gluon that can be
characterized by the two masses $M_G$ and $M_\Psi$ and the degree of
compositeness $g_c $. 

In this Section, we probe our set of \mch benchmark models with the suite
of $t\bar t$ resonance searches performed at the LHC8, as well
as planned searches at the LHC14.
As described in Section~\ref{sec:signatures},
there are three challenges in obtaining a signal at the LHC for heavy
gluons in the \mch scenario: 
\begin{itemize}
\item Because the top partners are already constrained to lie above
  800 GeV, heavy gluon production is suppressed due to smaller
  available phase space. 
\item The signal final-state is characterized by a top-quark pair and
  up to two extra massive gauge or Higgs bosons. This difference with
  respect to the pure $t\bar t$ final state systematically shifts
  events to values of $m_{t\bar t}^{true}$ much smaller than the
  $G$ mass. 
\item The large multiplicity of available channels implies that in
  most of the cases the heavy gluon is a rather broad resonance, which
  leads to a strong departure from the narrow width approximation. 
\end{itemize}

It is clear from the discussion above that the $t\bar t$ resonance
searches are likely to have much reduced sensitivity to the heavy
gluon in the $\rm MCH4_5$ model. However, a quantitative statement
requires a more careful treatment as
reconstructing the heavy gluon mass involves measuring top quarks in a
wide range of transverse momentum with techniques that were optimized
for the $t\bar t$ final-state hypothesis. In this situation, there are
two issues which might arise: 
\begin{itemize}
\item The sensitivity to any resonance search depends on the number of
  events available for analysis, which is affected by the overall
  efficiency of each selection. The precise interplay of the
  resolved and boosted analyses is likely to be highly process
  dependent and requires a dedicated study. 
\item
The additional particles in the final state beyond $t\bar t$ can have
a significant impact on the ability to experimentally resolve the
underlying parton-level distributions of the top kinematic
observables. The issue of resolution is inseparable from the genuine
new physics effects that distort the invariant mass distributions of
top pairs, as a wrongly reconstructed top will lead to an incorrect
estimate of the kinematic properties of the truth-level objects. 
 \end{itemize}

With this in mind we are going to study whether and to what extent the
new decay topologies 
might imply any observable excess in physical distributions.
For the purpose of heavy gluon reconstruction, we find the analysis
performed by the ATLAS collaboration for their KK gluon benchmark 
points at the LHC8 run to be adequate to illustrate the general
situation without much loss of generality. 

While our focus in this Section is for most of the part on physical
distributions, we find it instructive to look back at the parton-level
truth information for events that pass our full set of
reconstructions, as this gives a feeling on how much mass degradation
is due to misreconstruction effects versus genuine new physics
effects. 
 In order to disentangle these two effects, we consider the $t\bar t$ invariant mass resolution defined as 
  \begin{equation}
\epsilon \equiv (m_{ t \bar t}^{\rm rec}- m_{t\bar t}^{\rm truth})/m_{t\bar t}^{\rm truth}
\end{equation}
with $m_{t\bar t}^{\rm reco}$ ($m_{t\bar t}^{\rm truth}$) being the invariant mass of the reconstructed (truth) top-quark pair. 
 Here, $\epsilon$ is computed on an event-by-event basis for events that pass the selection cuts. The resolution should vanish for a perfectly reconstructed top-quark pair.

In addition to the LHC8 searches, future data taking and enhanced
analyses at 14 TeV will greatly extend the expected coverage of
searches for heavy-colored particles decaying to $t\bar t$. Here, we
consider the impact of two different analyses. The first one is
similar to the lepton+jets search presented in
Ref.~\cite{ATL-PHYS-PUB-2013-003} by the ATLAS collaboration. We have
performed our own version of this analysis in a manner identical to
that employed above for the LHC8 by following ATLAS as closely as
possible. A full 
signal+background study using parton-level truth information is also 
provided as a limiting 
example of the expected reach at the 14 TeV LHC, independently of any
particular experiment 
or top reconstruction method. 

The second analysis attempts to reconstruct the $G$ mass by making use
of all jets in the final state. Among the main issues one might face
for such an analysis is the challenge of identifying and tagging many
jets in the very intense hadronic activity that this kind of events
produce.  As we focus here on the present experimental status of the
heavy gluon searches, we do not consider the reconstruction of the top
partners in full and leave a detailed analysis of these signatures for
future work. Rather, we content ourselves with pointing out a
simple-minded analysis to extend the reach of these searches, in the
case in which the heavy gluon is a relatively narrow resonance.

\subsection{Physical distributions}\label{sec:physical_distributions}

\begin{figure}[htbp]
\begin{center}
\begin{tabular}{cc}
\includegraphics[width=3.2in]{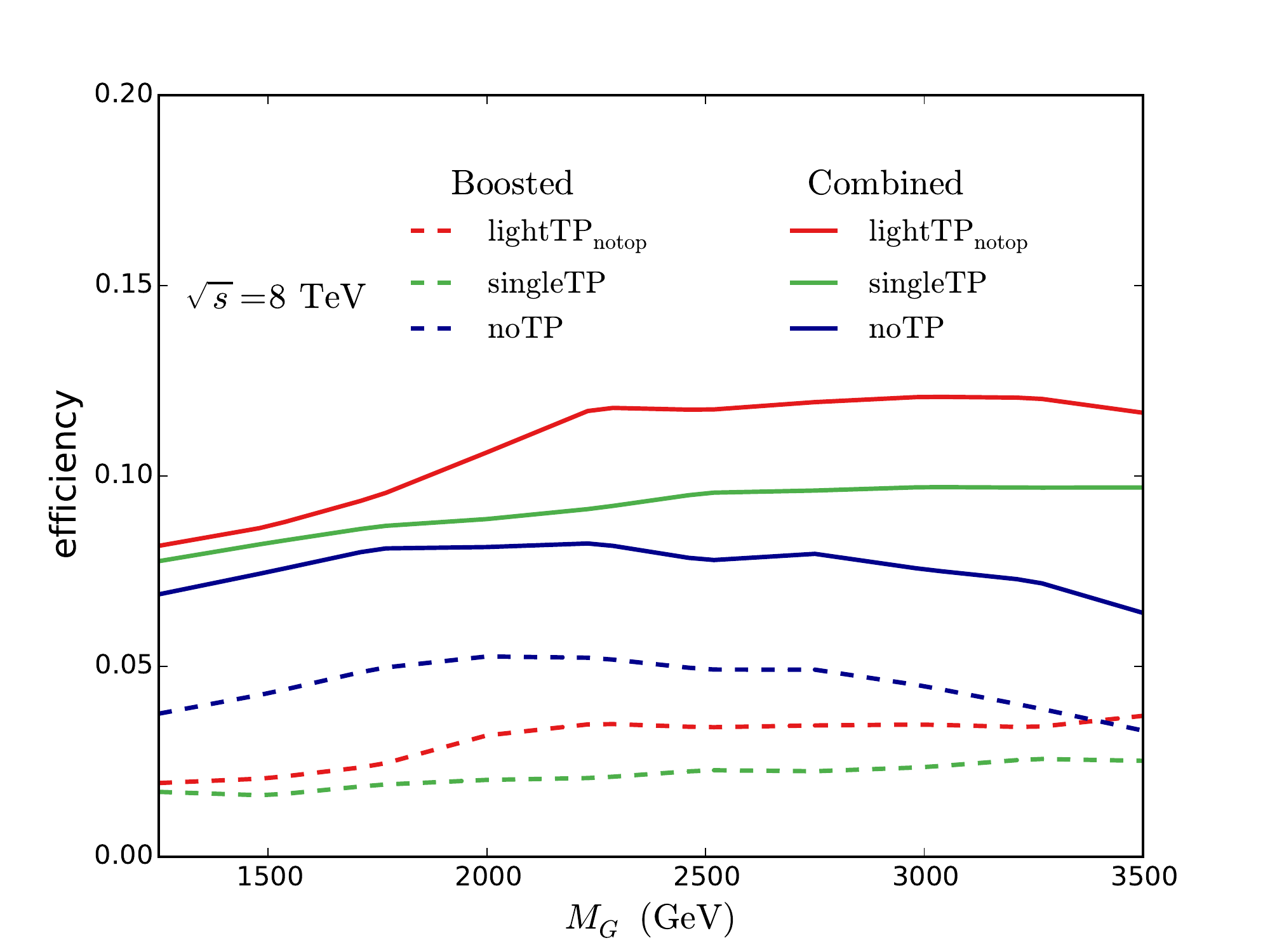} & \includegraphics[width=3.2in]{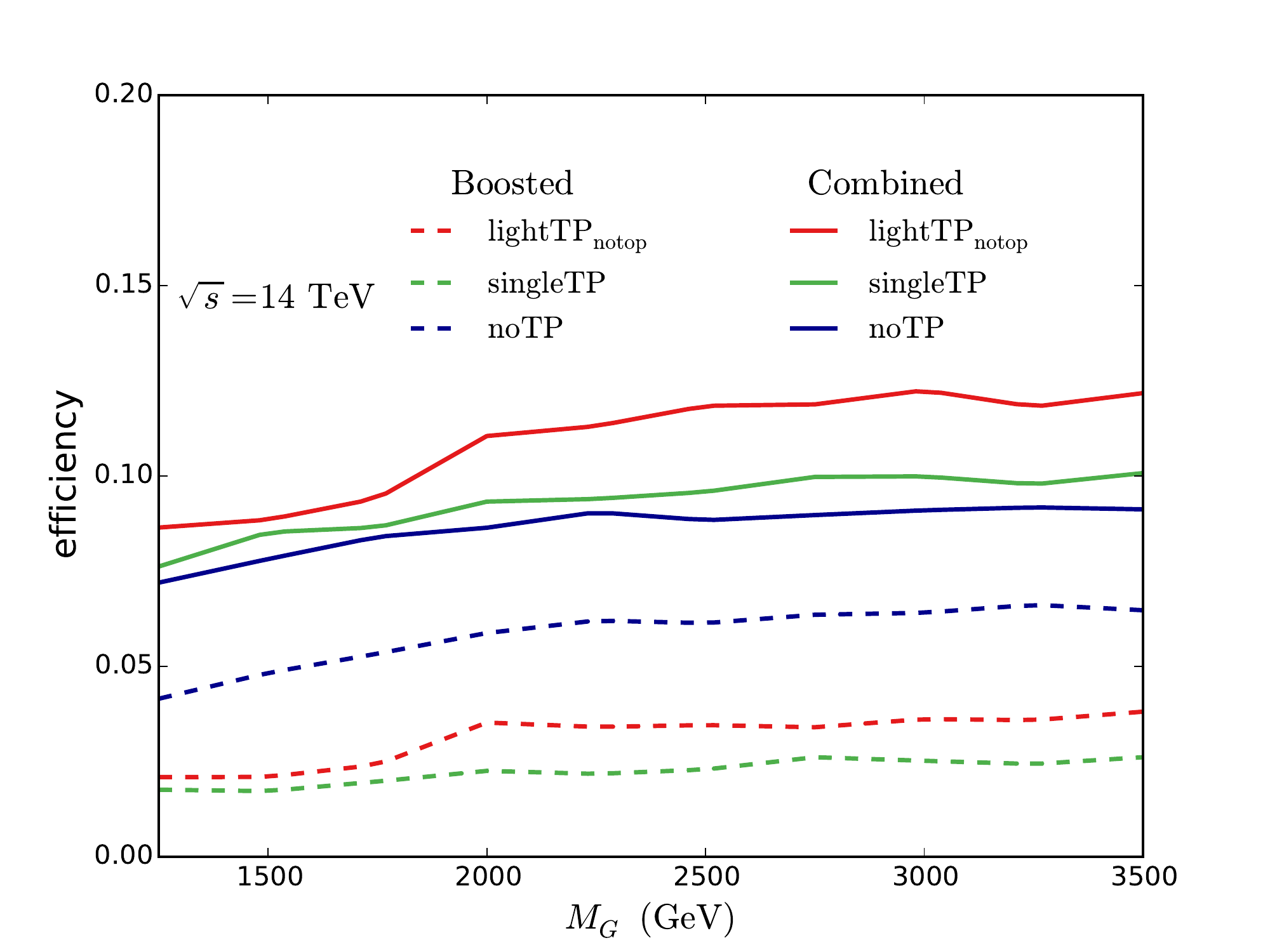} 
\end{tabular}
\caption{ The selection efficiency as a function of $M_G$. Dashed lines show the boosted selection and solid lines the total selection efficiency. The plots correspond to $M_\Psi = 1\TeV.$ }
\label{fig:eff}
\end{center}
\end{figure}

To get an idea of how the new topologies affect the event selection,
we study the performance of the selection cuts used by
ATLAS. Figure~\ref{fig:eff} shows the efficiency for the Boosted and
the final (Boosted+Resolved) selections on various models as a
function of the heavy gluon invariant mass. Here we define the efficiency
for the TP models as the passing fraction of reconstructed $t\bar t$
events in the lepton+jets final state.  As expected, the Boosted
selection is less efficient in the case with new channels than in the
noTP model, because tops from top partner decays tend to have lower
transverse momentum than those coming directly from heavy gluon
decays, and the Boosted selection is optimized to tag tops at very
high $p_T$. 
 For the combined selection, however, the TP topologies can have
 higher pass rates than the noTP model.  This is because the Resolved
 reconstruction is designed to reconstruct events with low $m_{t\bar
   t}$ and, hence, allows for more efficient reconstruction of the TP
 events.~\footnote{ However, the Resolved selection does not include a
   minimum quality cut on the $\chi^2$ distribution and as such does
   not guarantee that the invariant mass of the two reconstructed top
   quarks will reflect the invariant mass of the underlying
   truth-level tops with enough precision. }  This result suggests
 that a 
 purely boosted search for $t\bar t$ final states might not be the
 optimal strategy for models with top partners. Resolved analysis
 might still provide relevant information even as we increase the
 reach in mass.

\begin{figure}[htbp]
\begin{center}
\begin{tabular}{cc}
\includegraphics[width=3.2in]{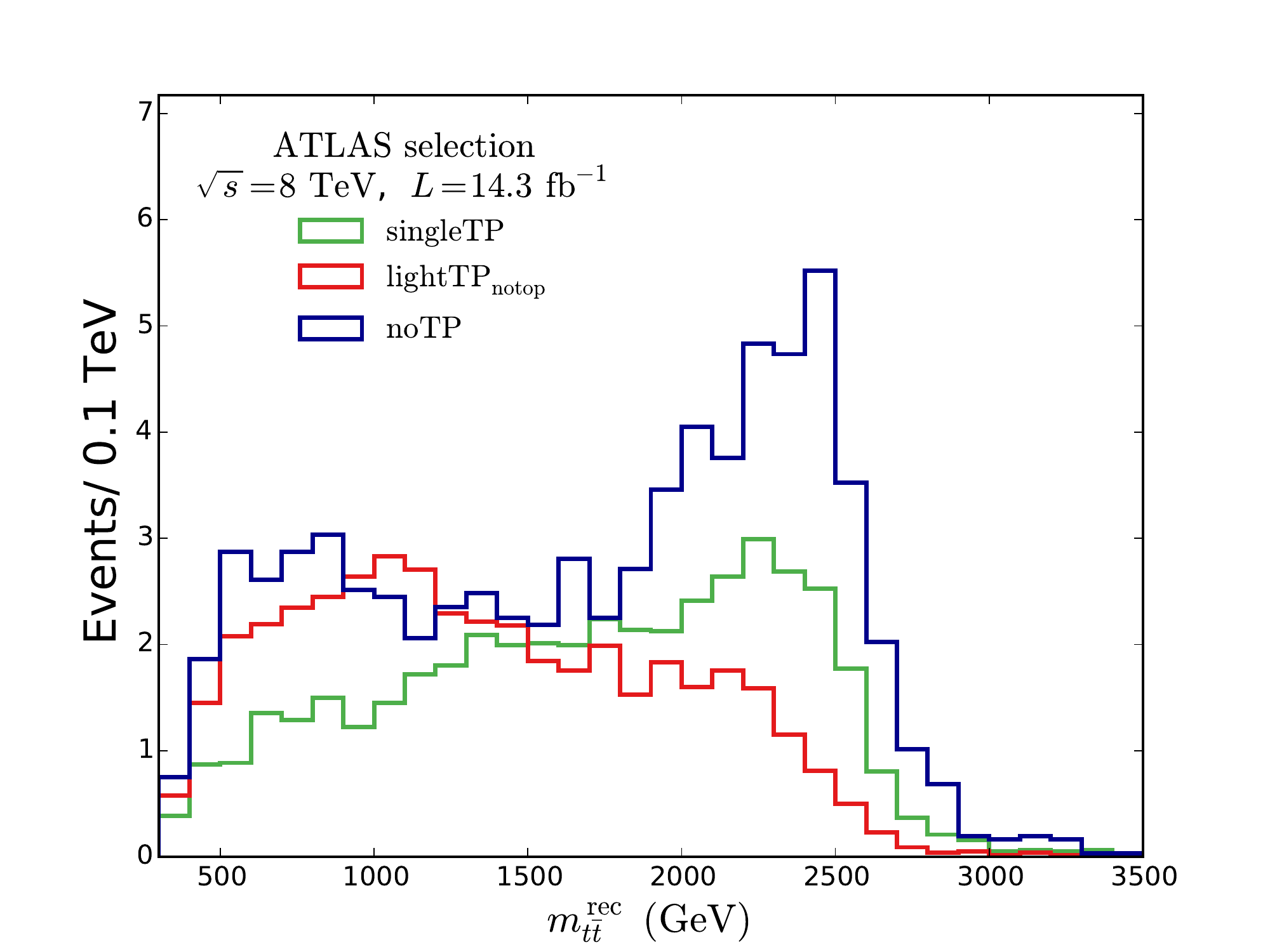}  &
\includegraphics[width=3.2in]{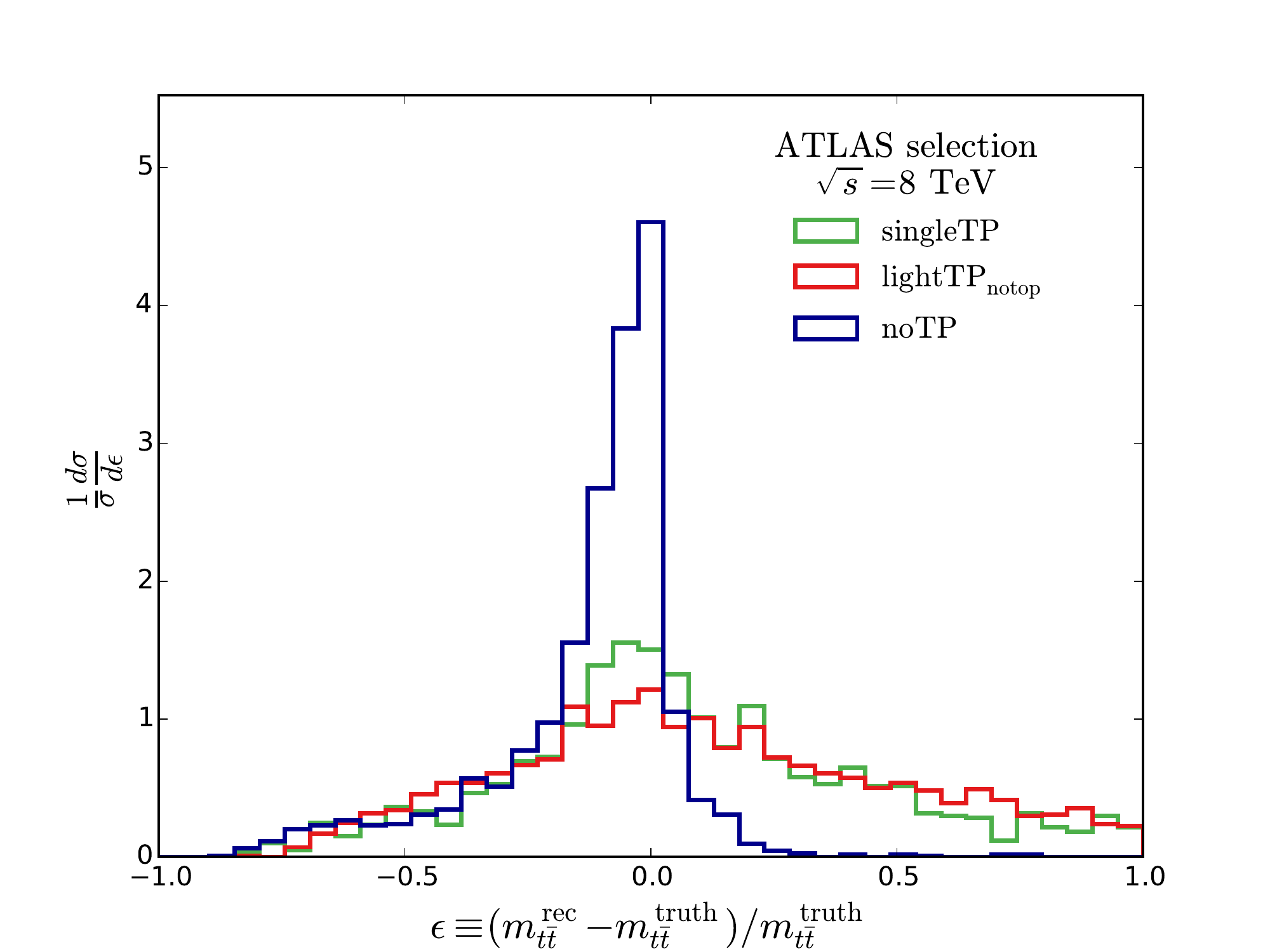}  \\
\includegraphics[width=3.2in]{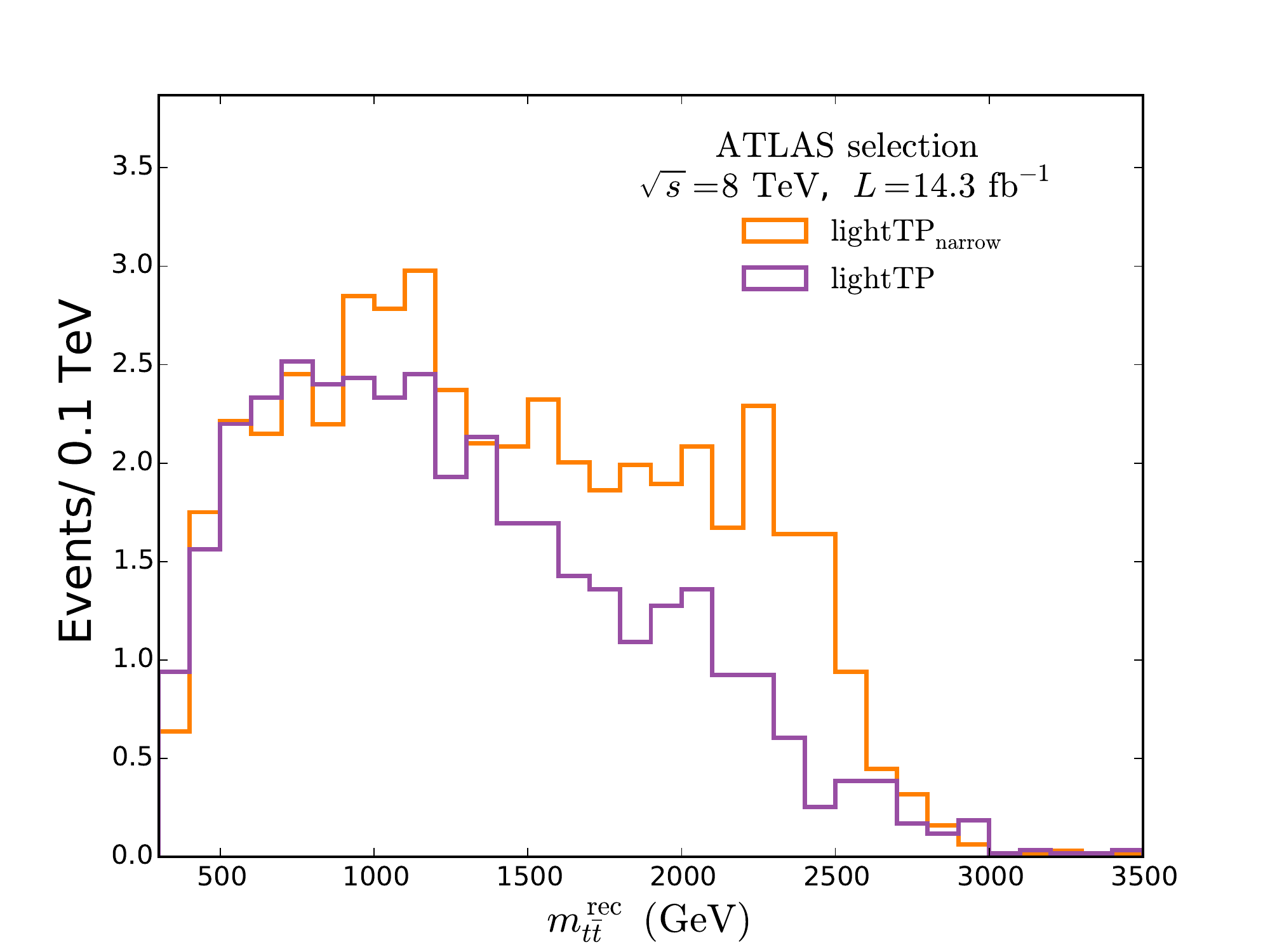}  &
\includegraphics[width=3.2in]{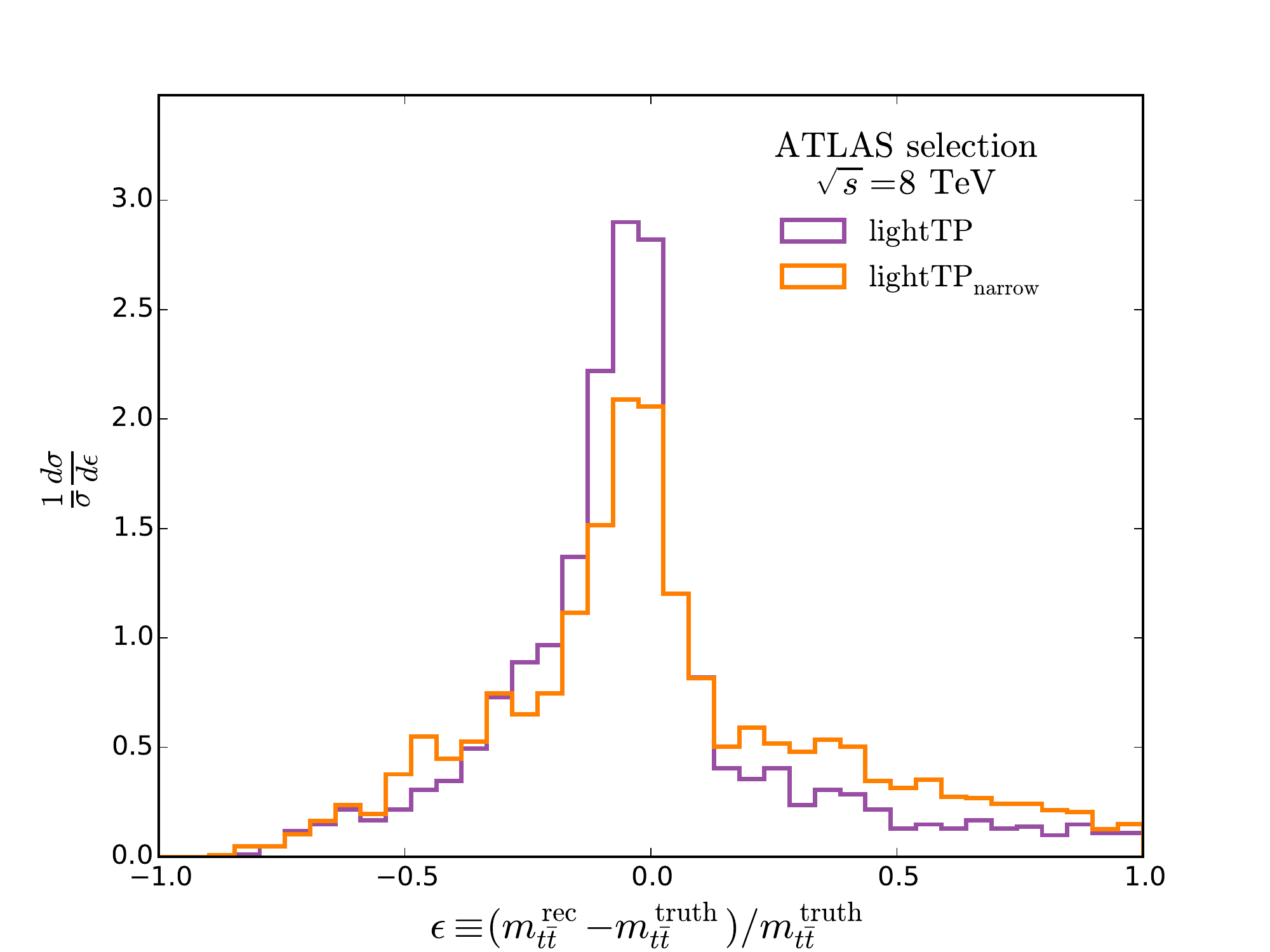}  \\
\end{tabular}
\caption{ Left panels: The reconstructed $t\bar t$ invariant mass distribution in signal events with 14.3 $\rm fb^{-1}$ at $\sqrt{s} = 8 \TeV$. Right panels: di-top invariant mass resolution. $m_{t\bar t}^{\rm truth}$ denotes the true (partonic) invariant mass of the top-quark pairs, while $m_{t\bar t}^{\rm rec}$  is the reconstructed invariant mass of the top-quark pairs using the ATLAS analysis. The plots correspond to  $M_G= 2.5$ and $M_\Psi = 1\TeV$ for the five benchmark models.   }
\label{fig:mtt_part_vs_rec:ltp_vs_htp}
\end{center}
\end{figure}

In Figure~\ref{fig:mtt_part_vs_rec:ltp_vs_htp} (left), we show the
distributions of reconstructed  $t\bar t$ invariant mass, $m_{t\bar
  t}^{\rm rec}$, for the signal only and $M_G=2.5$ TeV. The results
are normalized to the LO production cross section obtained from
MadGraph and 14.3 $\rm fb^{-1}$ of integrated luminosity.  While the
noTP distribution shows the resonance structure corresponding to a
heavy gluon with $M_G=2.5 \TeV$,   the invariant mass distributions
for the models with new channels are significantly
distorted. Interestingly enough, although the reconstructed
distributions are consistent 
with the parton-level distributions in Section~\ref{sec:signatures},
we see that the number of events at high $t\bar t$ invariant mass is
increased in the TP distributions compared to the parton-level
results. Indeed, the $t\bar t$ analysis
appears to preferentially select $t\bar t$ solutions which reconstruct
to form a larger invariant mass, which manifests itself in events with
higher $m_{t\bar t}$ in comparison with
Figure~\ref{fig:mtt_part_8tev}. In the resolved analysis, this is
because the extra particles in the final state can  lead to a wrong
assumption on the reconstructed top when determining the $b\mbox{-}W$
combination from genuine tops. In the boosted analysis, one fat jet
can over-collect jets from the extra bosons which are not far away
from the tops. Due to loose mass constraints employed in the top quark
tagging algorithm, misidentification of this fat jet as $t$  leads to
an incorrect estimate of the kinematics of the top pairs. These fake
candidates are legitimate as far as the semileptonic top-pair
selection is concerned and in fact, even though 
they do not faithfully reflect the
underlying truth-level tops from the hard process that originated
them, they allow us to partially recover some of the signal. The use
of more efficient top taggers would however go in the opposite
direction, reconstructing distributions that are more similar to the
partonic ones and therefore worsening the reach in the heavy gluon
searches.
To make the comparison with our parton-level results more evident, in
Figure~\ref{fig:mtt_part_vs_rec:ltp_vs_htp} (right), we show
distributions of the resolution $\epsilon$ for the same benchmarks. 
We can see that the resolution for the TP models (those with light top
partners) is significantly
worse than that for the noTP model. 

\begin{figure}[htbp]
\begin{center}
\begin{tabular}{cc}
\includegraphics[width=3.2in]{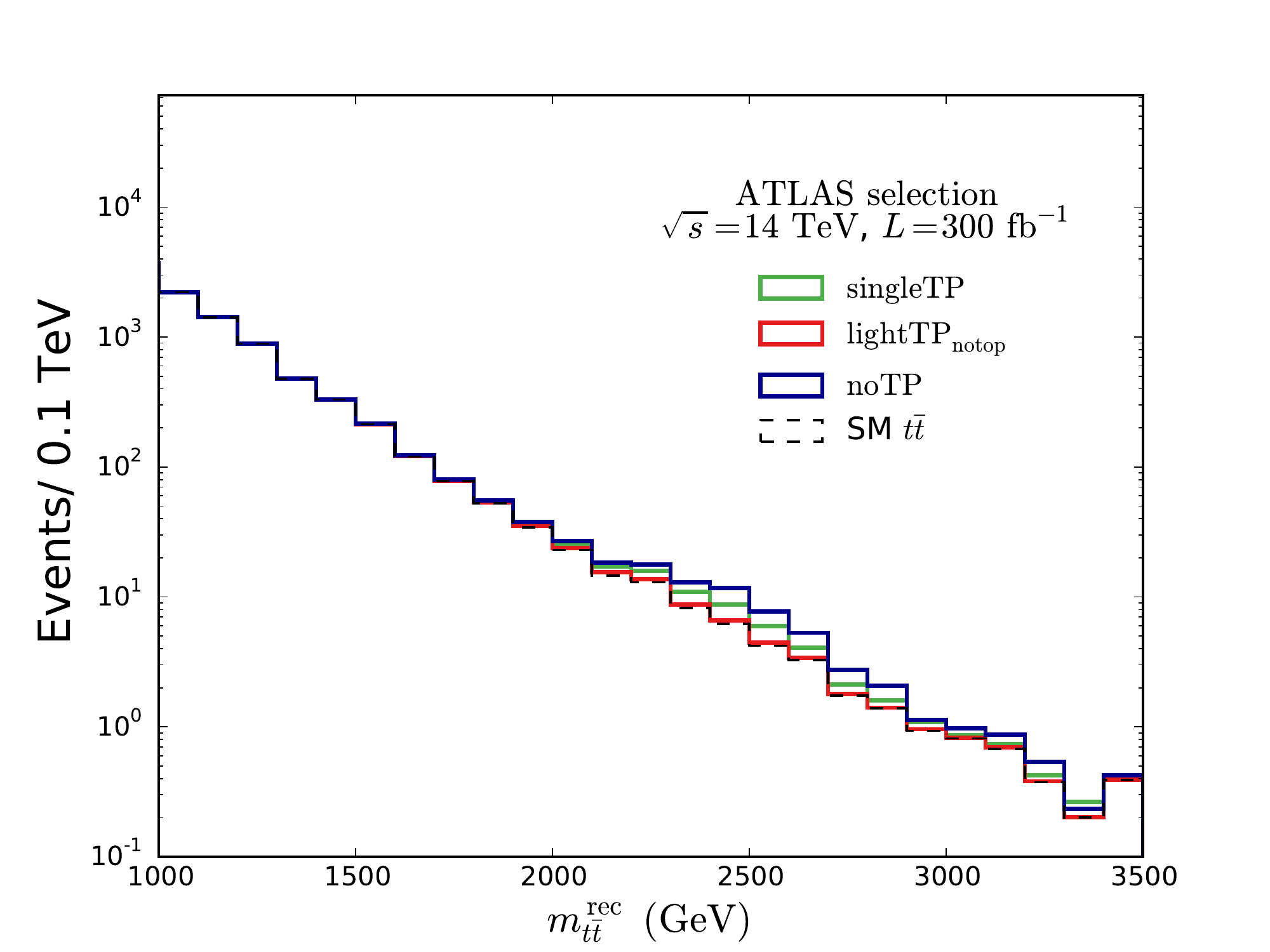} &
\includegraphics[width=3.2in]{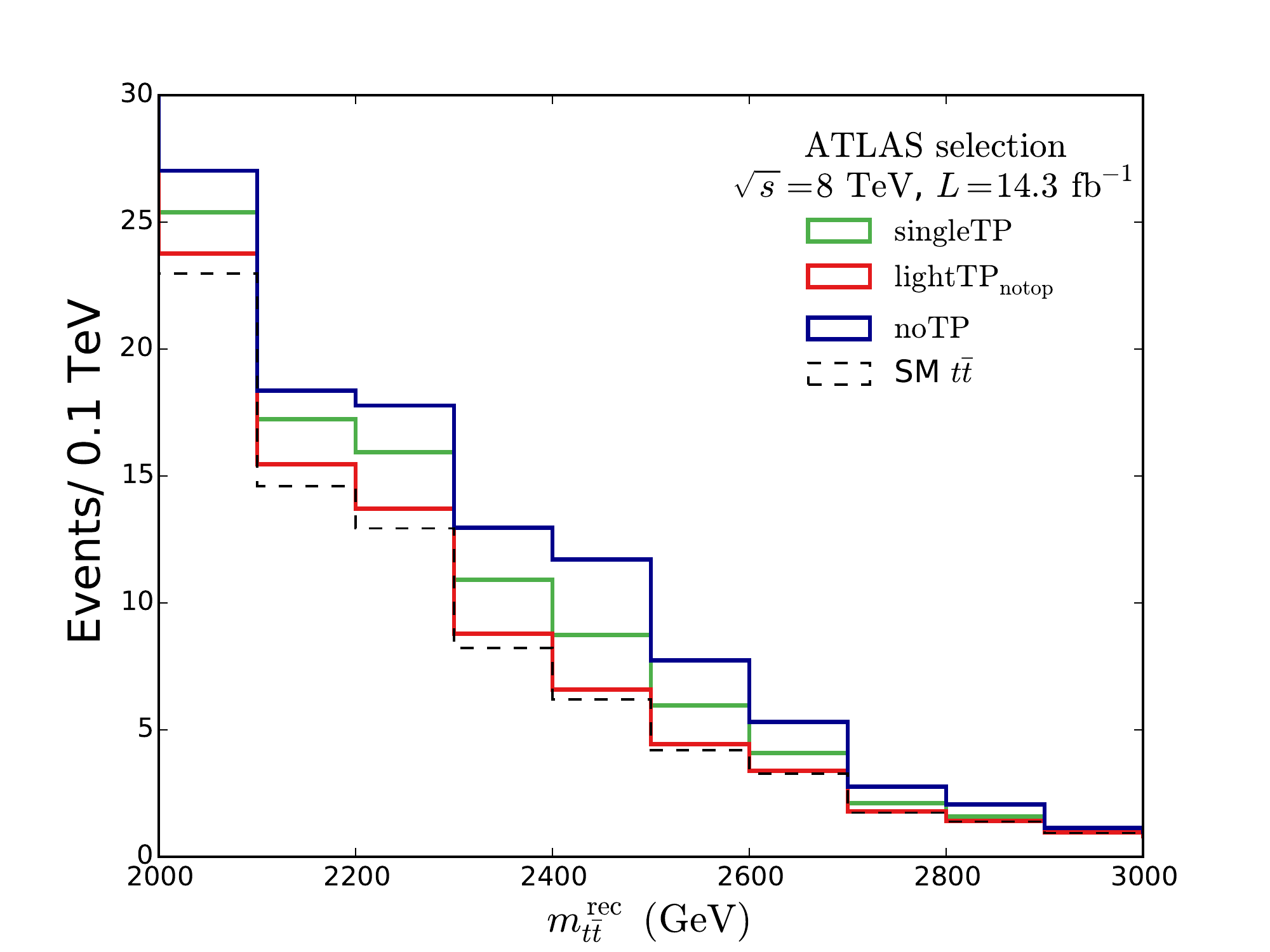} \\

\includegraphics[width=3.2in]{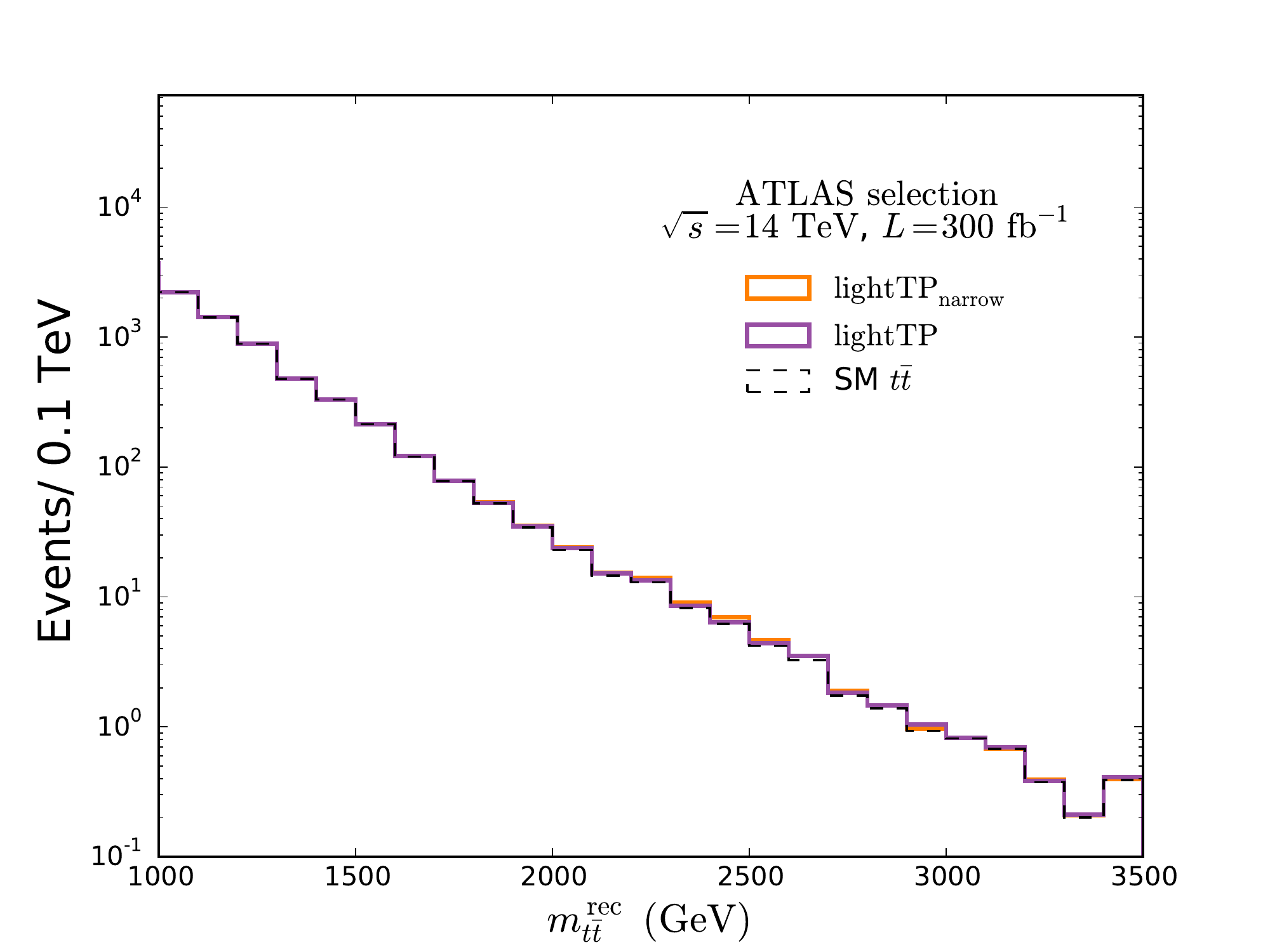} &
\includegraphics[width=3.2in]{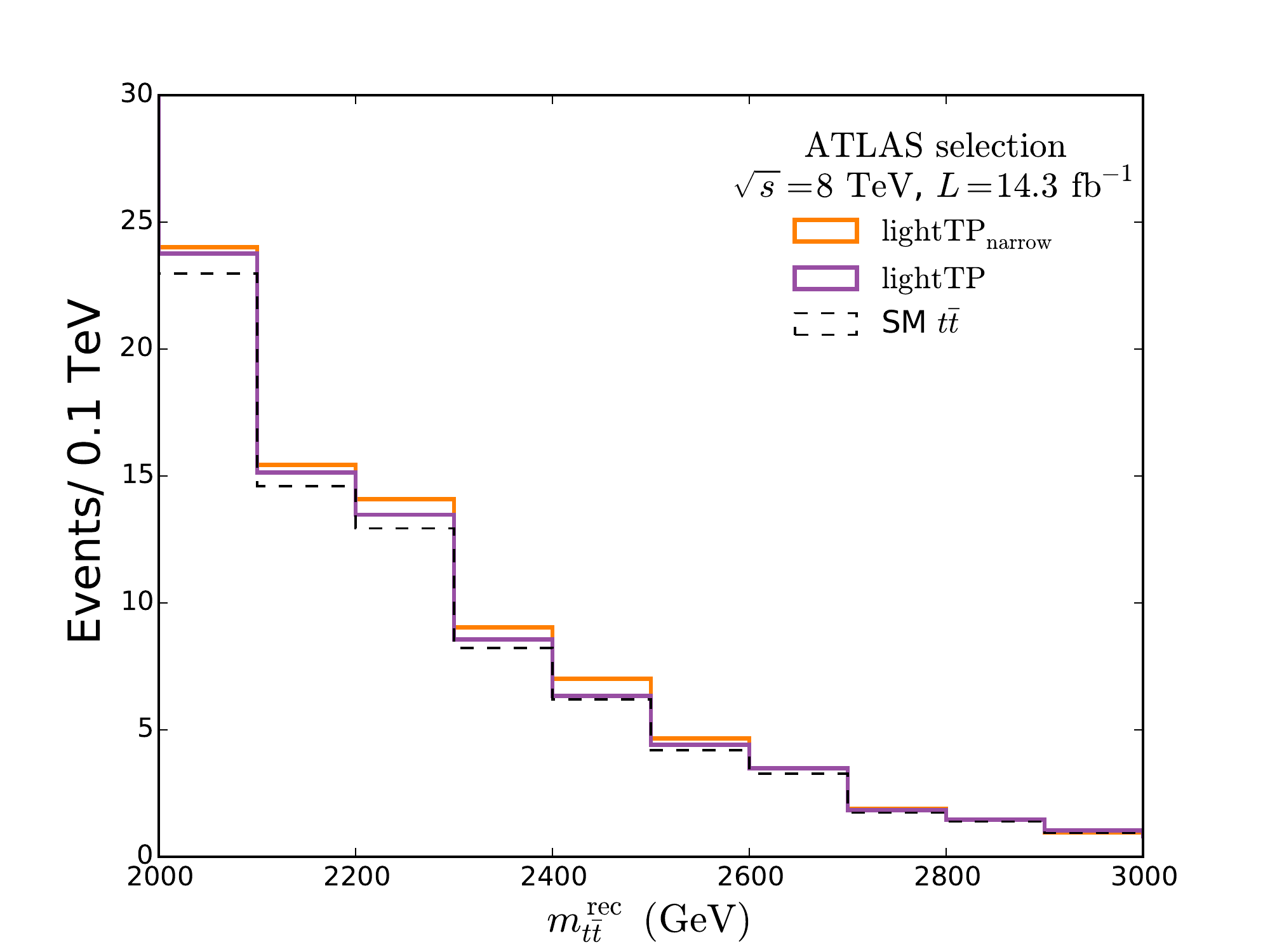} \\
\end{tabular}
\caption{Reconstructed $t\bar t$ invariant  mass distributions for heavy gluon production at the LHC with $\sqrt{s} = 8 \TeV$.   The solid histogram presents signal+background distribution, while the dashed histogram presents the $t\bar t$ SM background.  The right panels show the reconstructed invariant $t\bar t$ distribution focusing on the area near the would-be peak. The plots correspond to  $M_G= 2.5$ and $M_\Psi = 1\TeV$  for the five benchmark models.   Events are reconstructed using the ATLAS analysis by requiring the combined selection.  }
\label{fig:mtt_S_plus_B}
\end{center}
\end{figure}

Note that, although the used top taggers allow us to retain some of
the signals at invariant masses near the heavy gluon mass, the
distributions are still quite broad towards smaller values of the
invariant mass. 
Since the SM top pair production rate falls steeply as a function of
 the invariant mass, the net effect of the new TP topologies is to
 shift the contributions to a mass range where one would expect a
 higher SM $t\bar t$ background. It is therefore worth investigating
 whether the new processes could nevertheless lead to a potentially
 interesting signal.  In the left panels of
 Fig.~\ref{fig:mtt_S_plus_B}, we plot the total (signal + background)
 $m_{t\bar t}^{\rm rec}$ distributions for the five different models
 and $M_G=2.5 \TeV$.  In the right panels of
 Fig.~\ref{fig:mtt_S_plus_B}, we focus on the area near the peak. As
 expected, we see that the generic form of the resonance in the noTP
 model is clearly visible with the peak located at  about $M_G$.
 However, for the \stp~benchmark point the bump is much less visible,
 and for the \notop~benchmark, the observability of the heavy gluon
 signal is severely diminished. A similar behavior is obtained for the
 lightTP and lightTP$_{\mathrm{narrow}}$ models.
Even though we only show results for a   
 single mass point, we found that these features of the $t\bar t$
 spectrum are generic for all mass points for which the decays into
 top partners are kinematically open. 
A priori, since 
the lepton+jets searches at 8 TeV are less effective
at reconstructing the heavy gluon in the \mch scenario compared with
the RS-like KK  gluon, we would  
expect the same features to also be present for a single
14 TeV analysis.  
Indeed we have checked that a similar behaviour is obtained at the
LHC14. We do not reproduce the precise plots here as they are
qualitatively equal to the ones at 8 TeV.
The result is that essentially all models with top partners result in a much
distorted $t\bar t$ spectrum with a peak shifted 
to lower values of reconstructed $m_{t\bar t}$. As a consequence, the
signal in these models is much less visible over the continuum SM
$t\bar t$ 
than in the noTP case.

\subsection{Expected limits on heavy gluon mass}

\begin{figure}[htb]
\begin{center}
\begin{tabular}{cc}

\includegraphics[width=3.2in]{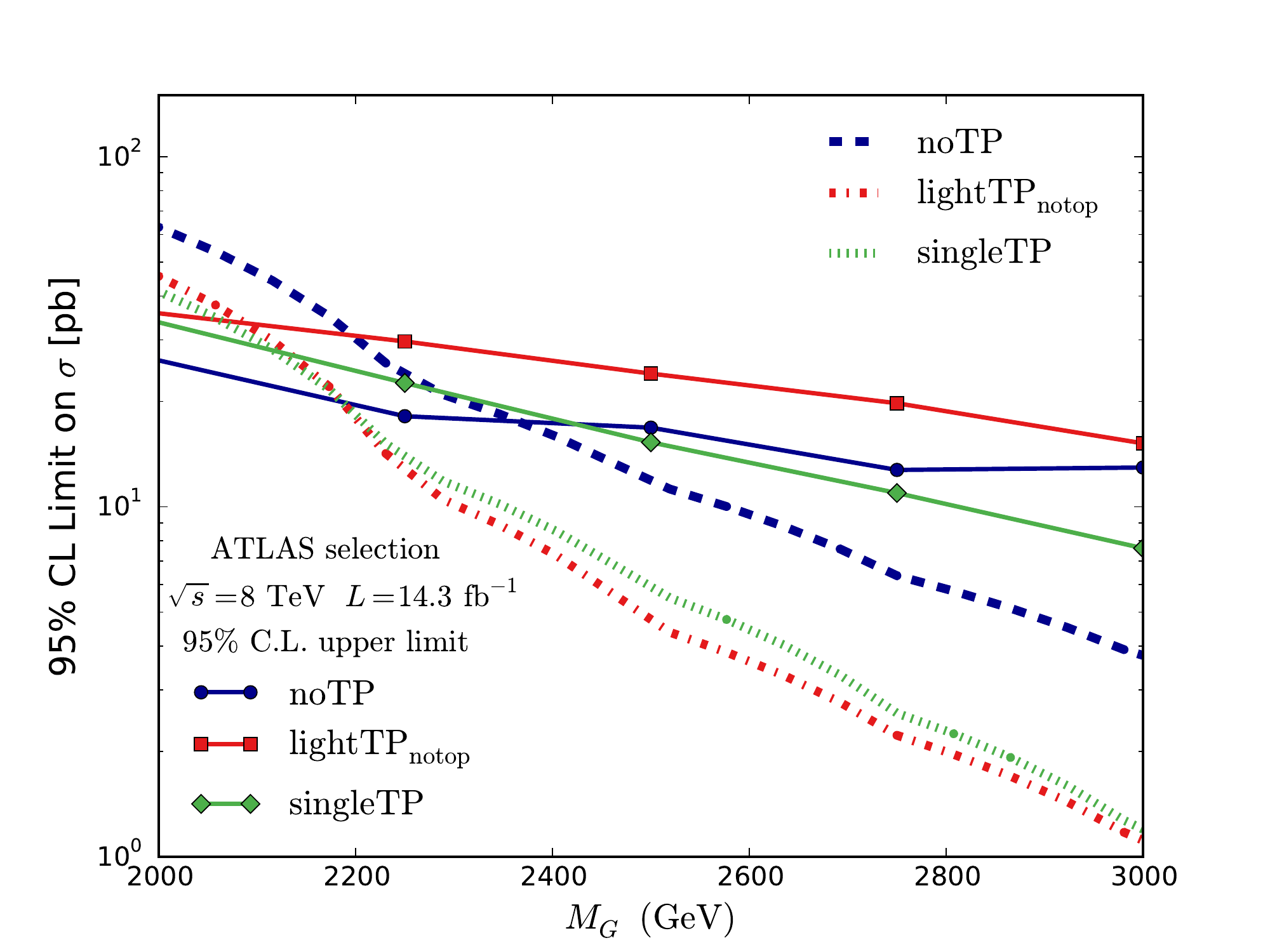} &

\includegraphics[width=3.2in]{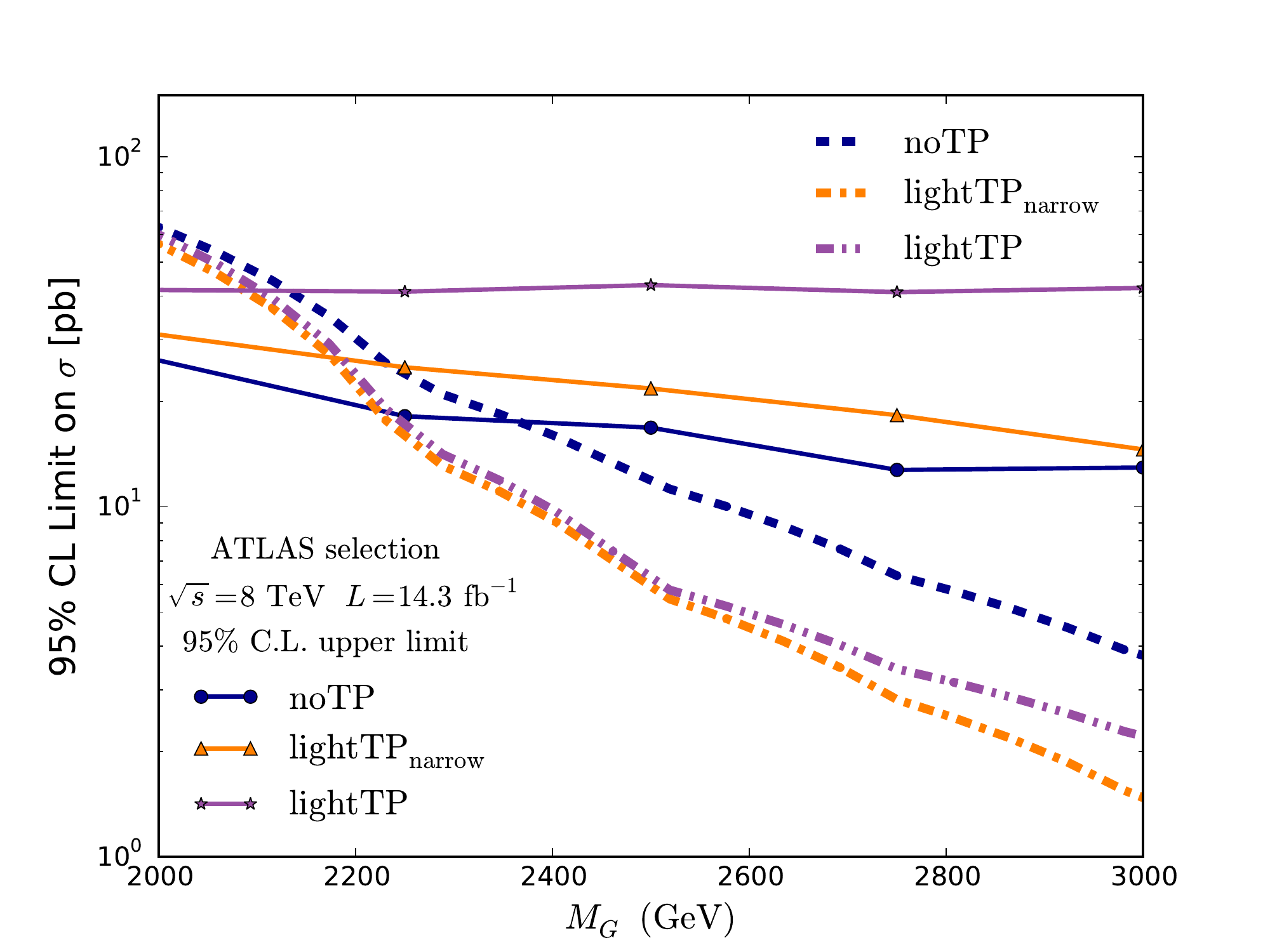} \\

\includegraphics[width=3.2in]{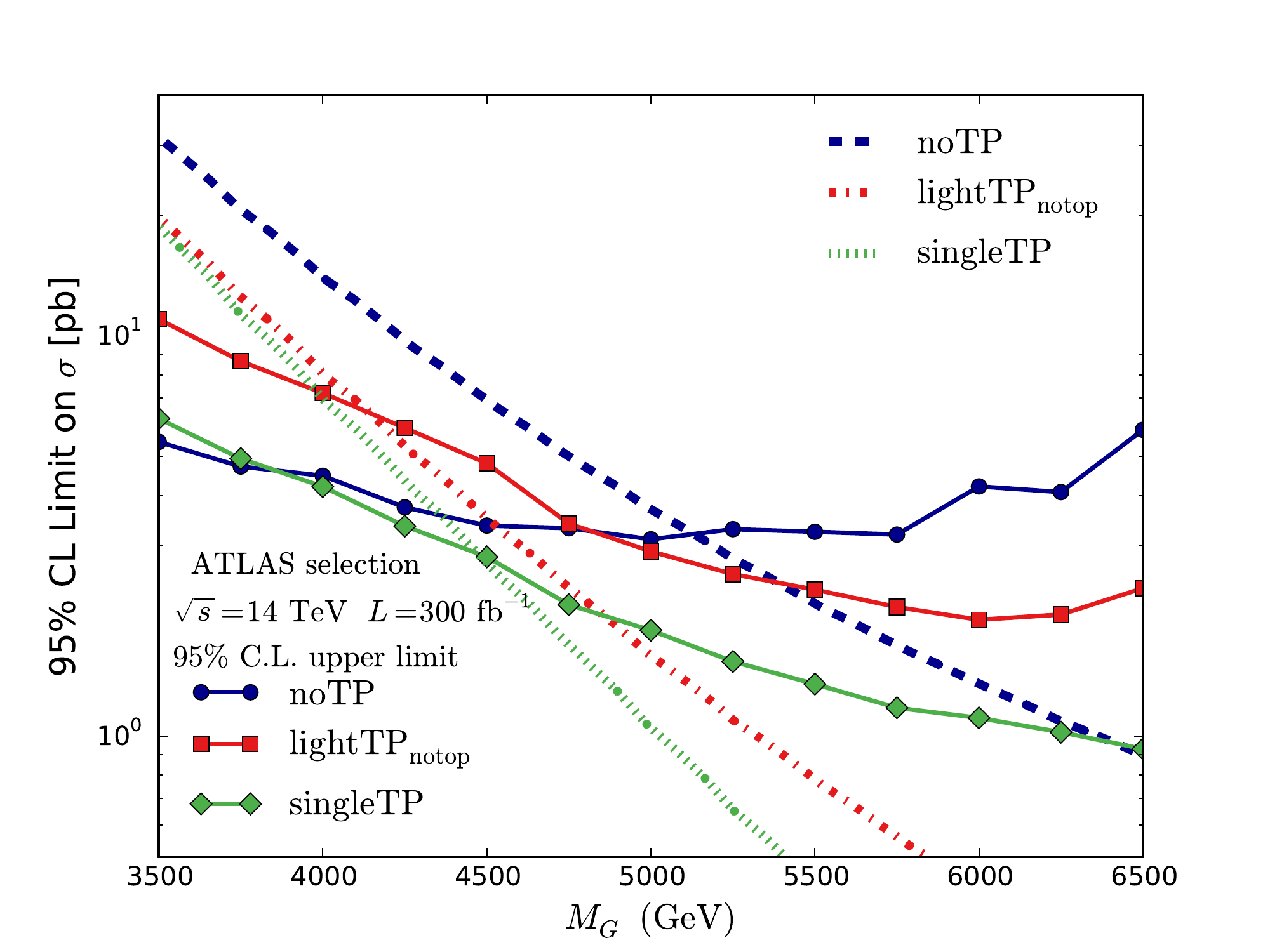} &

\includegraphics[width=3.2in]{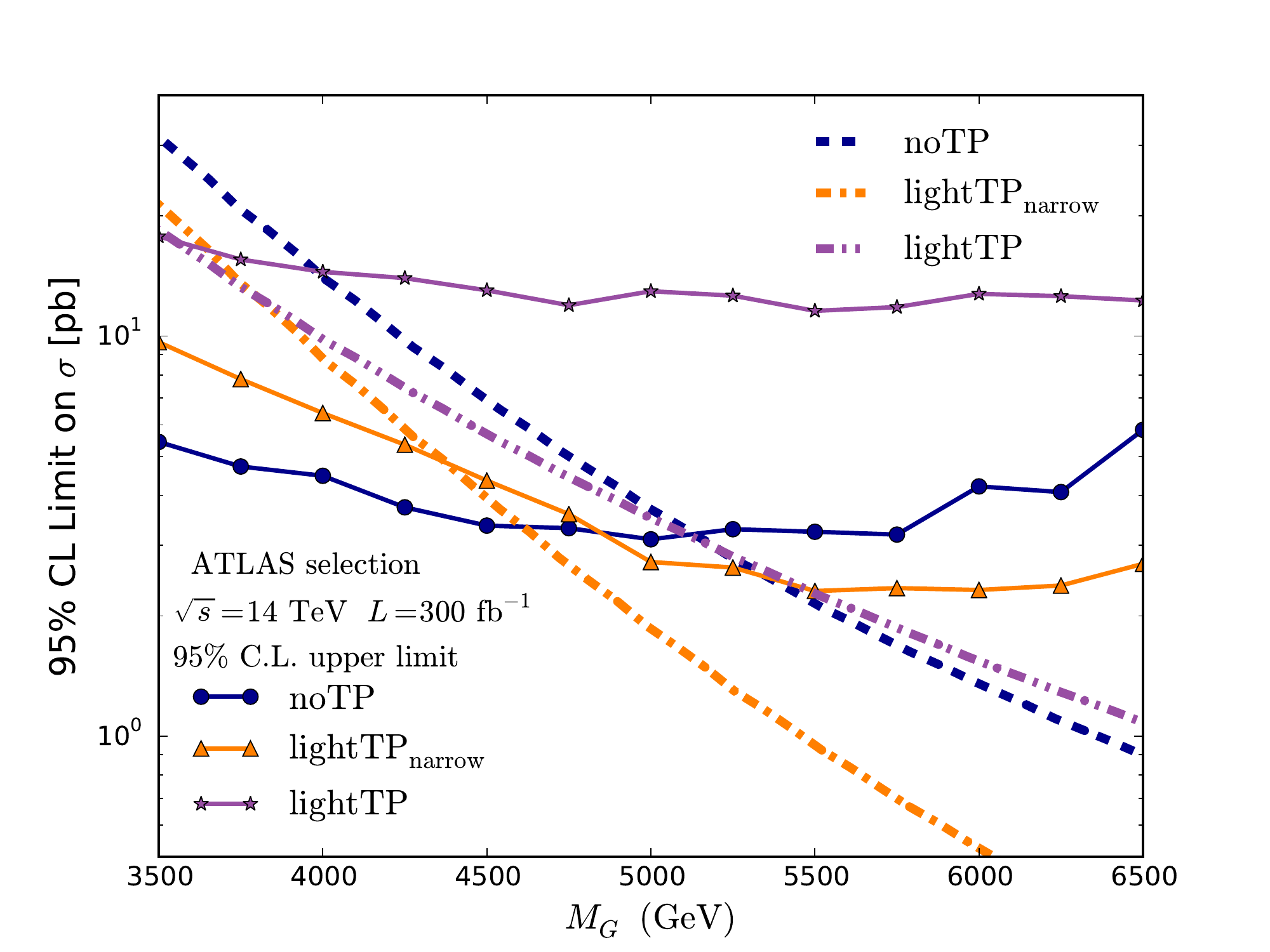}

\end{tabular}
\caption{The $95 \%$ confidence level limit (solid lines) on the heavy gluon production cross section as a function of $M_G$ is given for the five benchmark models. The theoretical cross sections are shown by the dashed lines. }
\label{fig:cls_mch4}
\end{center}
\end{figure}

The above results suggest that the $t\bar t$ resonance searches
should indeed have reduced sensitivity to exclude certain regions of the
\mch parameter space. We use a Bayesian statistical method to extract
the $95\%$ C.L. upper limits on the $pp\to G$ cross section as described
 in Section~\ref{sec:statistical_procedure}. For each $G$ mass point and
 benchmark model, we generate $10^4$ signal events using our MadGraph+Pythia
chain and scale them to an integrated luminosity of $14.3 \,\rm fb^{-1}$
 for $\sqrt{s} = 8 \TeV$ and  $300 \,\rm fb^{-1}$ for $\sqrt{s} = 14 \TeV$.
For each of the mass points investigated, we apply our reconstruction
in order to obtain a value for $\sigma \times BR \times \epsilon$
to be used in the Bayesian analysis. The results are shown by solid lines in
 Figure~\ref{fig:cls_mch4}. The heavy gluon production cross sections
 are indicated 
by the dashed lines in
Figure~\ref{fig:cls_mch4}. 

The performance of the $t\bar t$ search depends strongly on the final state topology.
For $M_G \lesssim 2 \TeV$, decays into top partners are closed, and we
return to the noTP case. For $M_G \gtrsim 2 \TeV$, 
the bounds are weakened considerably
because 1) the decay channels into top partners open up,
and this tends to dilute the signal over the SM $t\bar t$ background, and 2) the production cross section is reduced due
to suppressed phase-space and departure from the narrow-width approximation.
With $14.3 \,\rm fb^{-1}$ of 8 TeV data, the bounds for realistic heavy gluons can be up to $\sim 400 \GeV$ less stringent than the bound for the heavy gluon in the noTP model.
The increase in the center of mass energy to 14 TeV and  integrated luminosity
to $300 \,\rm fb^{-1}$ makes a significant impact on the overall
\mch model coverage. However, the differences between the models
 are still significant: realistic heavy gluons can be up
to 1.5 TeV lighter than the heavy gluon in the scenario
 considered by the experimental searches.

\begin{figure}[htb]
\begin{center}
\begin{tabular}{cc}
\includegraphics[width=3.2in]{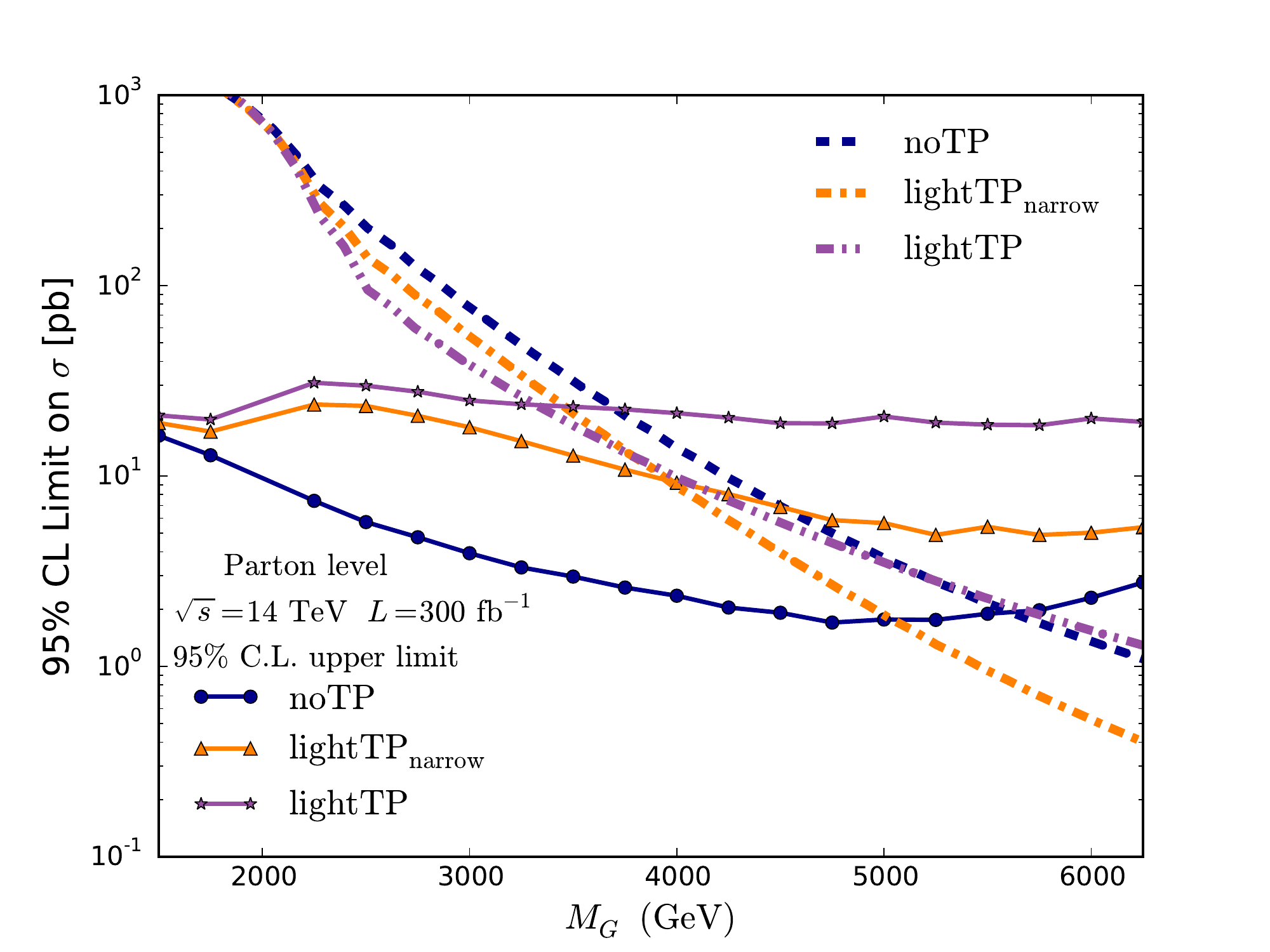} &

\includegraphics[width=3.2in]{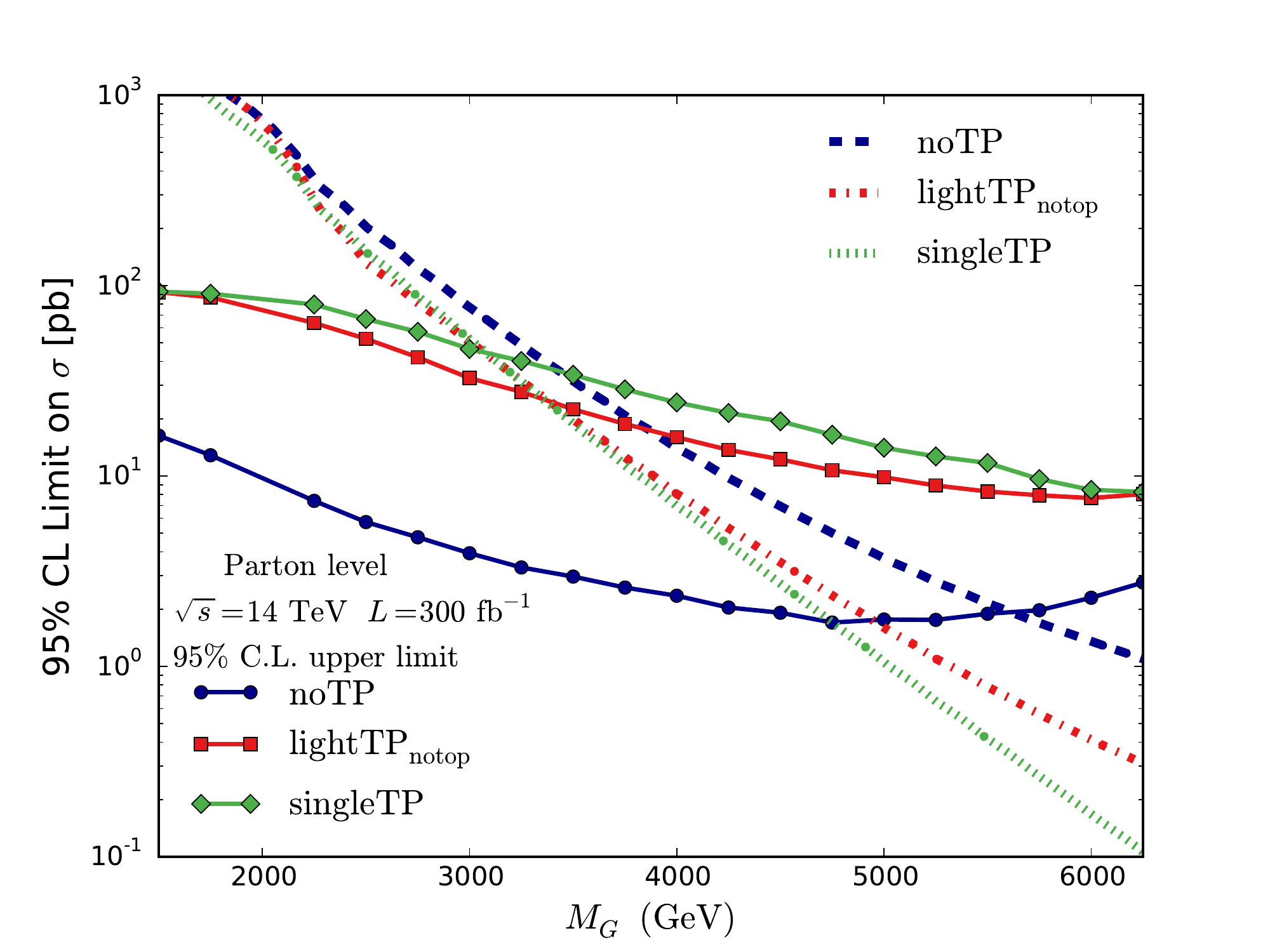}

\end{tabular}
\caption{The $95 \%$ confidence level limit (solid lines) on the heavy
  gluon production cross section as a function of $M_G$ is given for
  the five benchmark models using the parton-level information. 
  The theoretical cross sections are shown
  by the dashed lines.} 
\label{fig:cls_mch4_14tev_part}
\end{center}
\end{figure}

As we have already mentioned, future analyses -specifically those
making use of a dedicated top tagging tool- might make things
worse. An increase in the top reconstruction efficiency would make the
distributions more similar to those found using truth-level tops,
where the differences between the models were more significant.  A
parton level study, where tops are reconstructed perfectly, can be
used to set worst-case scenario limits. In
Fig.~\ref{fig:cls_mch4_14tev_part} we show the expected bounds on the
cross section $\sigma$ as a function of $M_G$ at 14 TeV using only
parton-level truth information.  This example shows that the limit
could be weakened by up to 2 TeV  with 300 $\rm fb^{-1}$ at the
LHC14. 

 This makes clear that the LHC discovery potential  on heavy gluon resonances may be crippled by a limited choice of
 search regions that may render impractical those analyses based
on the $t\bar t$ hypothesis. Nevertheless, several properties of the
top partners set them apart from ordinary top jets, and dedicated
searches could very likely extract the signal from the SM background
as in
Ref.~\cite{Barcelo:2011vk,Barcelo:2011wu,Bini:2011zb,Carmona:2012jk,Chala:2013ega}. Below we point out some features of the signal
that could be targeted by experiments in order to increase the
sensitivity. The discussion is concise and qualitative; a more
quantitative study is beyond the scope of this paper and will be
attempted elsewhere.

\subsection{Recovering the signal}

With such low signal sensitivity it is interesting to attempt a more
tailored search that would have an increased sensitivity. We try to be as
inclusive as possible and determine whether a search that was not
optimized to look for secondary resonances (the top partners) could
still discover the heavy gluon.  For the the mass of the top partners
considered in this paper,  the extra bosons are typically boosted and
their decay products are very collimated and can be caught within a
fat jet of large radius.  Therefore, we slightly change the analysis
by including the jets resulting from the hadronic decays of the extra
bosons in the top partner cascade decay. Using the same selection cuts
as above but without discarding the sub-leading  fat jets with $p_T >
200 \GeV$ we simply construct the invariant mass of the $t\bar t$
system and any sub-leading fat jet found by the following procedure.
The ATLAS $t\bar t$ reconstruction is first used to
select the event and reduce every final state to the ditop topology. Next,
we find all anti-$k_T$ $R=1.0$ jets, $j_i$, which are well separated from all
the jets that are part of the reconstructed $t\bar t$ system, 
\begin{equation}\label{eq:selection_G_reco}
\begin{array}{r@{}l}
{\rm Boosted~ selection:} \,\,&{} \Delta R (J, j_i) > 1.0, \,\,\,\,\Delta R (j_{\rm sel}, j_i) > 1.4\\ [6pt]
{\rm Resolved~ selection:} \,\,&{} \Delta R (j, j_i) > 1.4,
\end{array}
\end{equation}
where $j_{\rm sel}$($J$) are the selected small-radius (large-radius)
jet in the boosted reconstruction sample, and $j$ is any of the
selected small-radius jets in the resolved reconstruction sample. We
then combine the four momenta of these jets with the four momenta of
the reconstructed leptonically and hadronically decaying
top quarks. The candidate invariant mass, $M_G^{\rm reco}$,
is computed from the four momenta of all physics objects in the event,  
\begin{equation}
m_{G}^{\rm reco} = \left( p_{t_1} + p_{t_2} +\sum_{i \notin t_1,  t_2} p_i \right)^2,
\end{equation}
where $p_{t_i}, i=1,2$ are the four-momenta of the top candidates,
$p_i$ are the four-momenta of the $j_i$ and the sum runs over the
$j_i$ which satisfy  Eq.~(\ref{eq:selection_G_reco}). 

\begin{figure}[htb]
\begin{center}
\begin{tabular}{cc}
\includegraphics[width=3.2in]{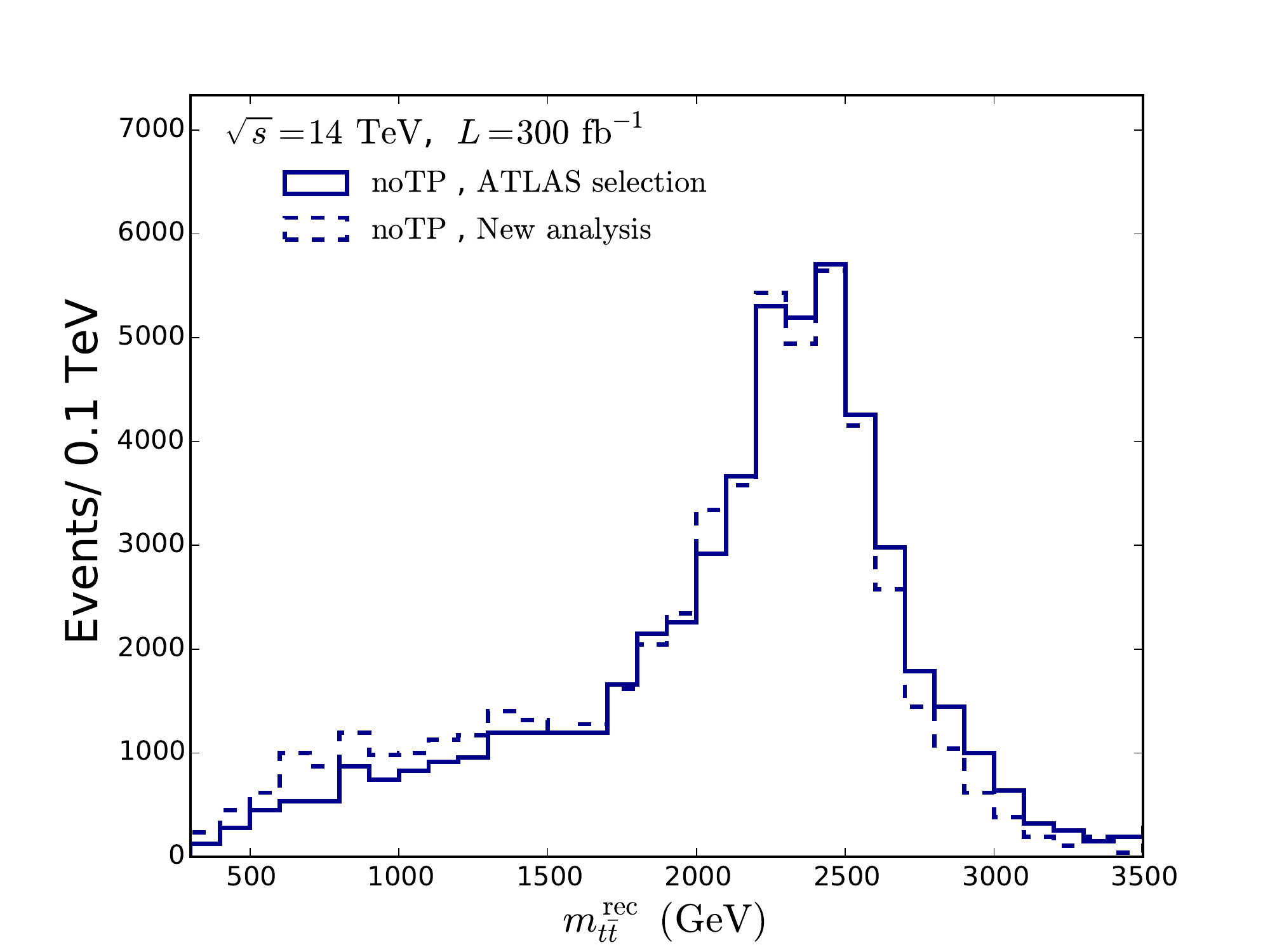} &
\includegraphics[width=3.2in]{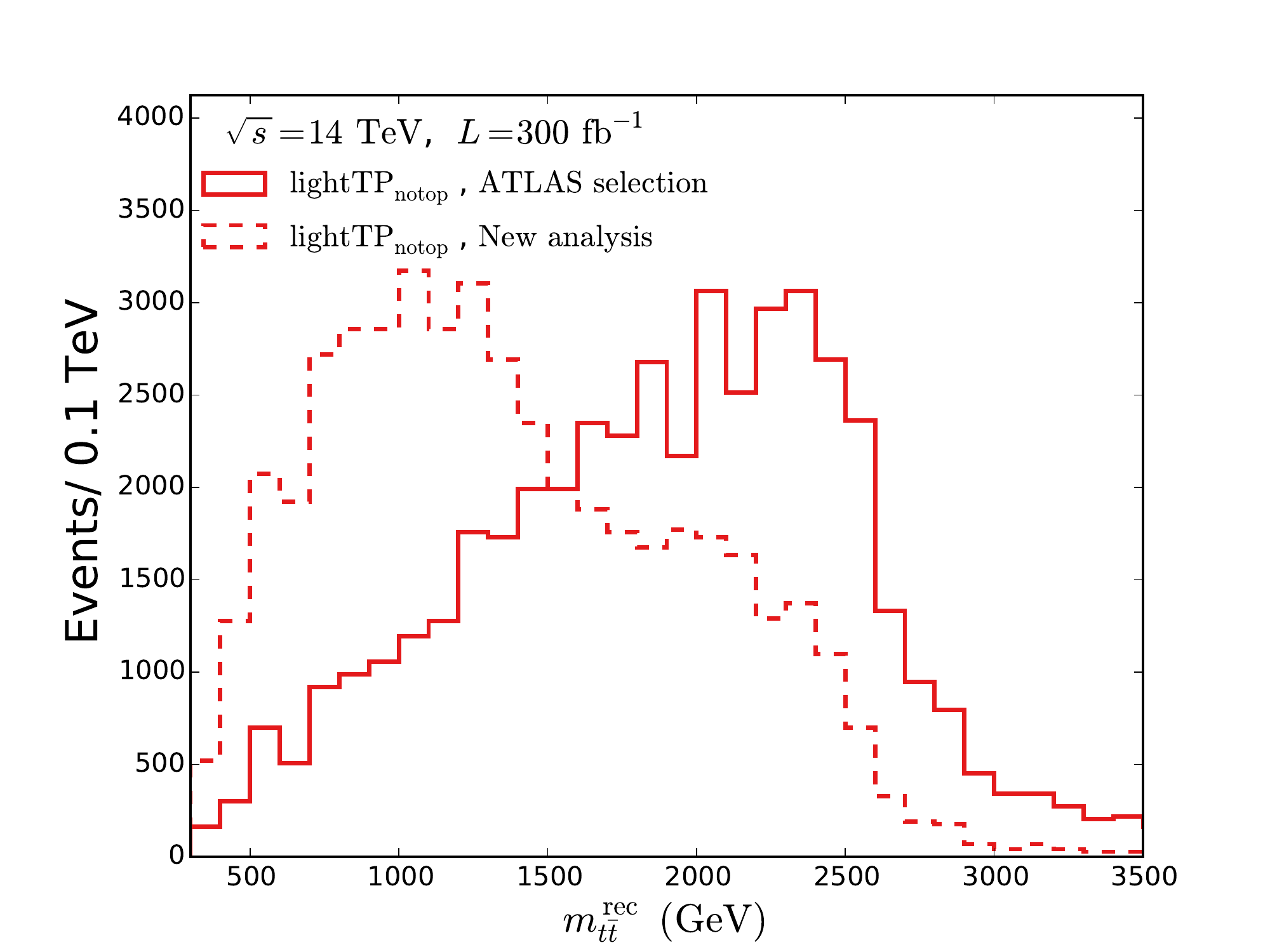} 

\end{tabular}
\caption{Reconstructed $G$ invariant mass distributions for noTP model (left) and \notop~model (right).   The plots correspond to  $M_G= 2.5 \TeV $ and $M_\Psi = 1\TeV$  in the ${\rm MCH}4_5$ model  with gc4 couplings.  }
\label{fig:modified_reco_14tev}
\end{center}
\end{figure}

Figure~\ref{fig:modified_reco_14tev} shows the distributions of the
reconstructed invariant $G$-mass using both methods; we show the
distributions for the noTP model (red) and the
lightTP$_{\mathrm{narrow}}$ model (blue) 
and $M_G=2.5 \TeV$. Also for comparison we show in dashed lines 
the distributions
using the ATLAS analysis.  These histograms show that, as expected,
the noTP model is not sensitive to the different reconstruction
methods but the lightTP ones can greatly benefit from the
modified analysis. Indeed, this new analysis provides a 
more efficient method to reconstruct the invariant mass of the heavy
gluon by recovering some of the heavy vector and Higgs
bosons coming from the top partners. Furthermore, we can use this
behaviour in the invariant mass to search for narrow resonances in the
background from continuum top partner pair production. 

\begin{figure}[htb]
\begin{center}
\includegraphics[width=4.in]{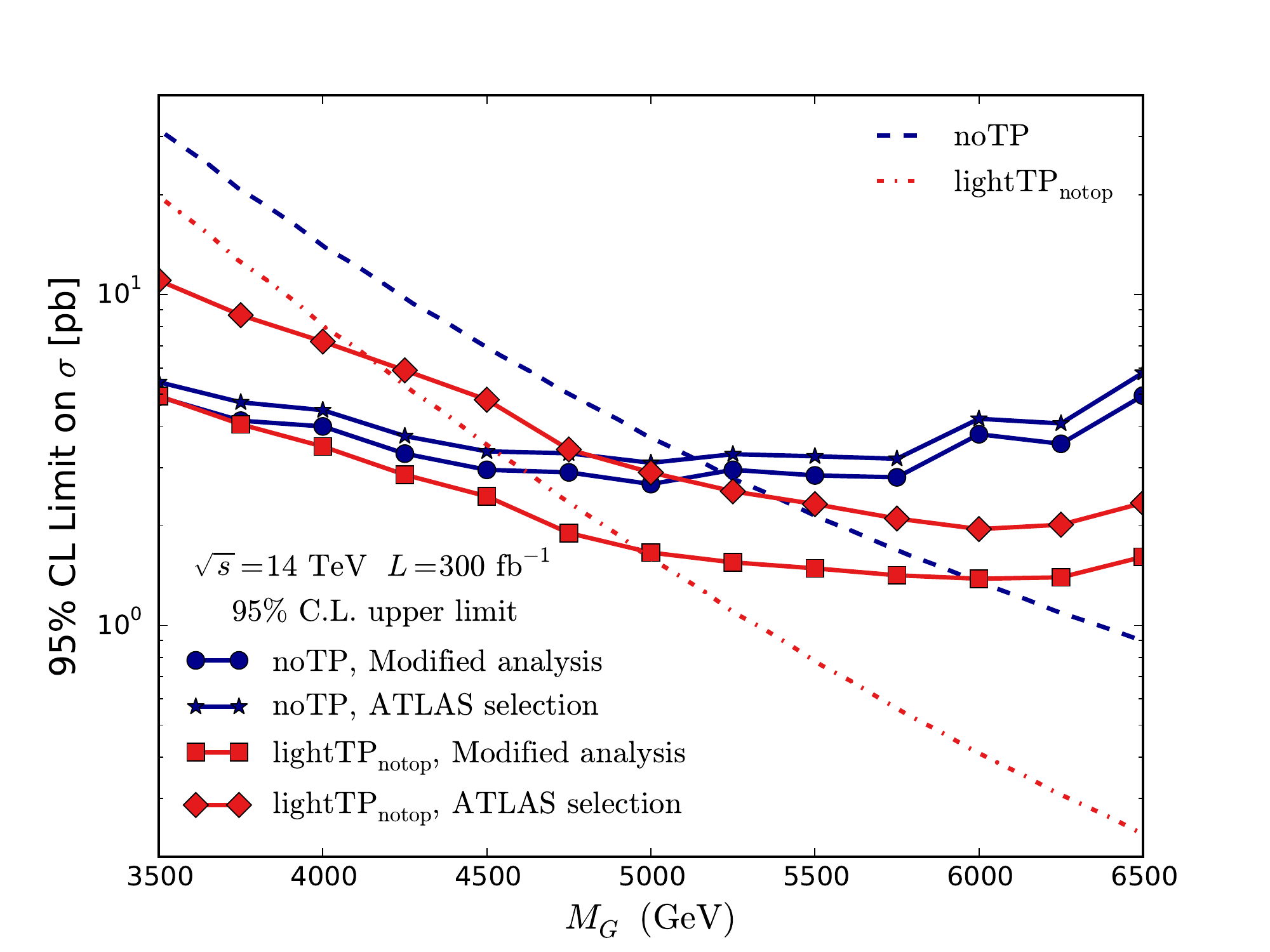} 
\caption{The $95 \%$ confidence level limit (solid lines) on the heavy
  gluon production cross section as a function of $M_G$ according to
  the ATLAS lepton+jets analysis and our modified analysis. 
The theoretical cross sections are shown by the dashed lines. }
\label{fig:cls_reco_14tev}
\end{center}
\end{figure}

 In Figure~\ref{fig:cls_reco_14tev} we present the results for
 $\sqrt{s} = 14 \TeV$ and a total integrated luminosity of 
$300 ~\rm fb^{-1}$. We show the 95 $\%$ C.L. upper limit on
 the heavy gluon production cross section as a function of $M_G$
 according to the ATLAS lepton+jets analysis for the noTP sample
 (blue) and for the \notop~sample (red). We also show in the same
 Figure for comparison the bound using
 the modified lepton+jets analysis for the same samples. As expected,
 no significant difference is seen on the noTP sample when using the
 modified analysis. The results presented in
 Figure~\ref{fig:cls_reco_14tev} allow us to conclude that a sizeable
 improvement in the bound can be achieved by including the extra
 hadronic activity in $t\bar t$ resonance searches. Furthermore it is
 clear that further improvement can be attained by performing
 full-fledged di-top-partner resonance searches, especially for the
 case of narrow resonances.

\section{Conclusions and outlook}\label{sec:conclusions}

Natural composite Higgs models generically contain both fermionic and vector
resonances. 
The former, called top partners, are
expected to couple strongly to the electroweak boson and gluon partners, and
this expectation is quantitatively confirmed by arguments based on
holography.  
The combination of bounds from direct and indirect searches and
naturalness arguments lead to the plausible and more viable scenario in
which there is a small hierarchy between the masses of the spin-half
and spin-one 
resonances. 
This hierarchy suggests an important change in the
way we look for the vector resonances 
at the LHC experiments. In this work we have focused on searches for a
heavy gluon but similar arguments apply to electroweak vector
resonances. The present experimental analyses are
not geared for the case in which the heavy gluon decays with a large
decay ratio into top partners, as we have explicitly demonstrated
in this paper. The final effect on the heavy gluon exclusion limits is
model dependent but we find a qualitative decrease in the
experimental bounds on the heavy gluon mass in the ${\rm
  MCH4}_5$ model. This means that even the simplest composite Higgs
scenarios provide highly non-standard gluon partner signals that elude
existing search strategies aimed at the RS-like KK gluon, composite
Higgs models or their close variants. 

Finally, a few words on the interplay of our results with top partner direct searches is in order.
While near future single production searches can potentially rather
quickly reach the 2\,TeV scale (see {\it e.g.}~\cite{Backovic:2014uma}
using state of the art substructure
methods~\cite{Almeida:2010pa,Backovic:2013bga}) they are however
rather model dependent. On the other hand the more robust searches via
pair production are very limited in their reach  
(see {\it e.g.} recent discussion of
reach~\cite{WillocqTalk}).
Indeed, the models that we have considered
in this work are not excluded so far by the direct searches. 
Consequently, one may ask whether one can use the new
resonance-top-partners production as a discovery channel. The answer
should be in principle positive (see~\cite{Carena:2007tn} for a
detailed discussion of this point in the context of holographic
composite Higgs models). 
One can envision two ways to go about it, the first was already
described by us in the previous Section, namely, increasing the
sensitivity via cutting hard on the activity in the event, say by
looking at extra hadronic activity. Similarly one can look at extra
contributions to the transverse mass or energy from leptonic and
missing energy type of deposition. As this should be rather efficient
way to increase the signal-to-background ratio one might be able to
extend the pair production reach and to allow for an early discovery. 
Finally, if the vector resonance is narrow (as could happen in
composite models for the coloured resonances and especially for the
electroweak ones), then one may hope to
be able to significantly extend the reach to regions which are well
beyond that of the direct production regime.

\emph{Acknowledgements:} \\
JJ would like to thank the Weizmann theory group for their
hospitality during the initial stages of this project.
MC would like to thank the CERN Theory Group and the Institute for
Theoretical Physics at ETH Z\"urich for hospitality during the
completion of this project. 
JJ is supported by the project A.AC..FISI.771 PRIN 2010-2011.
MC and JS are supported by MINECO, under grant
numbers FPA2010-17915 and FPA2013-47836-C3-1/2-P, 
the FPU program (MC), by the European Commission through the contract
PITN-GA-2012-316704 (HIGGSTOOLS) and by Junta de Andaluc\'{\i}a grants
FQM 101 
and FQM 6552. 
GP is supported by the IRG, ISF, the Gruber award, and ERC-2013-CoG
grant (TOPCHARM \# 614794). Furthermore, a significant part of this
work was done when GP held a Staff position at CERN and JJ was a
postdoctoral fellow at the Weizmann Institute. 
Note added: while this paper was in its final phase of
preparations~\cite{Greco:2014aza} appeared that have some level of
overlap with the results 
presented by us.

\appendix

\section{Model description}

We provide in this Appendix a detailed description of the relevant
features of the model. Further information in terms of motivation and
extra collider implications can be found in the original
reference~\cite{DeSimone:2012fs}.
We assume the first two SM quark generations and the RH bottom quark
to be fully elementary and the RH top quark to be fully composite. The
third generation SM quark doublet is embedded in a $5$ of $SO(5)$
while the top partners are assumed to live in a $4$ of $SO(4)$. In the
basis we are considering these embeddings read
\begin{equation}
(Q^5_L)_I = \frac{1}{\sqrt{2}} 
\begin{pmatrix} 
\mathrm{i} b_L \\
 b_L \\
\mathrm{i} t_L \\
-t_L \\
0
\end{pmatrix},
\qquad
\Psi^i = \frac{1}{\sqrt{2}}
\begin{pmatrix}
\mathrm{i}(B-X_{5/3}) \\
B+X_{5/3} \\
\mathrm{i}(T+X_{2/3}) \\
-T+X_{2/3}
\end{pmatrix},
\end{equation}
where $I=1,\ldots,5$ and $i=1,\ldots,4$, respectively. 

The Lagrangian involving the fermions and gluons reads, in the
elementary-composite basis 
\begin{eqnarray}
\mathcal{L} &=& 
\bar{q}_L\mathrm{i}\cancel{D} q_L 
+\bar{t}_R\mathrm{i}\cancel{D} t_R
+\bar{\Psi}\mathrm{i}(\cancel{D}+\mathrm{i} \cancel{e}) \Psi
- M_\Psi \bar{\Psi} \Psi 
\nonumber \\
&+&
\Big[ \mathrm{i} c_1 (\bar{\Psi}_R)_i \gamma^\mu d^i_\mu t_R
+ y f (\bar{Q}^5_L)^I U_{Ii} \Psi^i_R 
+ y c_2 f (\bar{Q}^5_L)^I U_{I5} t_R 
+\mathrm{h.c.}
\Big]
\nonumber \\
&-&
\frac{1}{2} \mathrm{Tr}[G^e_{\mu \nu}]^2
-\frac{1}{2} \mathrm{Tr}[G^c_{\mu \nu}]^2
+\frac{1}{2} M_c^2 \left( G^c_\mu -\frac{g_e}{g_c} G^e_\mu\right)^2, 
\label{Lag:MCH45}
\end{eqnarray}
where the contribution from the Goldstone boson matrix $U$ and the $d$
symbol is given explicitly below.
In the Lagrangian above $f$ is the scale characterizing the strong
coupling scale while $y$ and $c_{1,2}$ are dimensionless parameters
expected to be of order one. The covariant derivatives read explicitly
\begin{eqnarray}
\mathrm{i}D_\mu q_L &=& 
\left(
\mathrm{i}\partial_\mu
+ g \frac{\sigma^i}{2} W^i_\mu + \frac{g^\prime}{6} B_\mu + g_e
G^e_\mu
\right)
q_L,
\nonumber \\
\mathrm{i}D_\mu t_R &=& 
\left(
\mathrm{i}\partial_\mu
+ \frac{2g^\prime}{3} B_\mu + g_c
G^c_\mu
\right)
t_R,
\nonumber \\
\mathrm{i}D_\mu \Psi &=& 
\left(
\mathrm{i}\partial_\mu
+ \frac{2g^\prime}{3} B_\mu + g_c
G^c_\mu
\right)
\Psi.
\end{eqnarray}
Note that the elementary (composite) gluon $G^e_\mu$ ($G^c_\mu$) 
couples only to the elementary (composite) quarks with coupling $g_e$
($g_c$). The mass matrix for the gluons can be diagonalized by means
of the following rotation
\begin{equation}
\begin{pmatrix}
G^e_\mu \\ G^c_\mu 
\end{pmatrix}
=
\begin{pmatrix}
\cos \theta_3 & -\sin \theta_3 \\
\sin \theta_3 & \cos \theta_3
\end{pmatrix}
\begin{pmatrix}
g_\mu \\ G_\mu
\end{pmatrix},
\end{equation}
with
\begin{equation}
\tan\theta_3=\frac{g_e}{g_c},
\end{equation}
so that we end up with the massless SM gluon, $g_\mu$, that couples
universally to all the quarks with coupling
\begin{equation}
g_s=g_c \sin\theta_3=g_e \cos\theta_3,
\end{equation}
and a heavy gluon, $G_\mu$, with mass 
\begin{equation}
M_G=\frac{M_c}{\cos \theta_3},
\end{equation}
and couplings to elementary and composite fields
\begin{eqnarray}
g_e G_\mu^e &=& - g_e \sin \theta_3 G_\mu + \ldots =
-\frac{g_s^2}{\sqrt{g_c^2-g_s^2}} G_\mu + \ldots, \\
g_c G_\mu^c &=&  g_c \cos \theta_3 G_\mu + \ldots =
\sqrt{g_c^2-g_s^2} G_\mu +   \ldots, 
\end{eqnarray}
respectively.

The remaining terms in the Lagrangian (\ref{Lag:MCH45}) read
\begin{eqnarray}
\mathrm{i}\bar{\Psi}_R^i \cancel{d}_i t_R &=&
\frac{g}{\sqrt{2}} s_h [(\bar{X}_{5/3})_R \cancel{W}^+ 
-\bar{B}_R \cancel{W}^- ]t_R 
\nonumber \\
&&-\frac{g}{2c_W} s_h [\bar{T}_R + (\bar{X}_{2/3})_R]\cancel{Z} t_R
+\mathrm{i} [(\bar{X}_{2/3})_R - \bar{T}_R]
\frac{\cancel{\partial}\rho}{f} t_R,,
\\
\bar{\Psi}\left(\frac{2g^\prime}{3} \cancel{B} - \cancel{e}
\right)\Psi
&=&
\frac{g}{c_W}\left(-\frac{1}{2}+\frac{s_W^2}{3}\right) 
\bar{B} \cancel{Z} B
+\frac{g}{c_W}\left(\frac{1}{2}-\frac{5s_W^2}{3}\right) 
\bar{X}_{5/3} \cancel{Z} X_{5/3}
\nonumber \\
&+&
\frac{g}{c_W}\left(\frac{1}{2}c_h-\frac{2s_W^2}{3}\right) 
\bar{T} \cancel{Z} T
+
\frac{g}{c_W}\left(-\frac{1}{2}c_h-\frac{2s_W^2}{3}\right) 
\bar{X}_{2/3} \cancel{Z} X_{2/3}
\nonumber \\
&+& \frac{g}{\sqrt{2}}
\Big\{ 
\bar{B} \cancel{W}^- \left[ c_{h/2}^2 T + s_{h/2}^2 X_{2/3}\right]
+\bar{X}_{5/3} \cancel{W}^+ \left[ s_{h/2}^2 T + c_{h/2}^2 X_{2/3}\right]
+\mathrm{h.c.} \Big\}
\nonumber \\
&+& \mbox{ photon couplings},
\\
(\bar{Q}^5_L)^I U_{Ii} \Psi^i_R &=&
 \bar{b}_L B_R +  \bar{t}_L 
\Big[ c_{h/2}^2 T_R  + s_{h/2}^2 (X_{2/3})_R \Big], 
\\
 (\bar{Q}^5_L)^I U_{I5} t_R &=&
-\frac{1}{\sqrt{2}} s_h \bar{t}_L t_R,
\end{eqnarray}
where we have denoted
\begin{equation}
s_x \equiv \sin \frac{x}{f}, \quad
c_x \equiv \cos \frac{x}{f}, 
\end{equation}
except for $s_W$ and $c_W$, which are the sine and cosine of the
Weinberg angle. $\rho$ is the physical Higgs boson and $h$ reads, in
the unitary gauge
\begin{equation}
h \equiv \langle h \rangle + \rho,
\end{equation}
with 
\begin{equation}
f s_{\langle h \rangle} = v \approx 246\mbox{ GeV}.
\end{equation}



\begin{thebibliography}{10}

\bibitem{PerezTalk}
G. Perez,
\newblock {Talk at the 2014 Zurich Pheno' Workshop, Jan/2014 }, 2014.

\bibitem{Chatrchyan:2013uxa}
  S.~Chatrchyan {\it et al.}  [CMS Collaboration],
  ``Inclusive search for a vector-like T quark with charge $\frac{2}{3}$ in pp collisions at $\sqrt{s}$ = 8 TeV,''
  Phys.\ Lett.\ B {\bf 729} (2014) 149
  [arXiv:1311.7667 [hep-ex]].

\bibitem{Aad:2014efa}
  G.~Aad {\it et al.}  [ATLAS Collaboration],
  ``Search for pair and single production of new heavy quarks that decay to a $Z$ boson and a third-generation quark in $pp$ collisions at $\sqrt{s}=8$ TeV with the ATLAS detector,''
  arXiv:1409.5500 [hep-ex].

\bibitem{TheATLAScollaboration:2013kha}
  The ATLAS collaboration,
  ``A search for $t\bar{t}$ resonances in the lepton plus jets final
  state with ATLAS using 14 inverse fb  of pp collisions at $\sqrt{s}=8$ TeV,''
  ATLAS-CONF-2013-052, ATLAS-COM-CONF-2013-052.


\bibitem{Chatrchyan:2013lca}
  S.~Chatrchyan {\it et al.}  [CMS Collaboration],
  ``Searches for new physics using the $t\bar{t}$ invariant mass
  distribution in pp collisions at $\sqrt{s}$ = 8 TeV,''
  Phys.\ Rev.\ Lett.\  {\bf 111} (2013) 21,  211804
   [Erratum-ibid.\  {\bf 112} (2014) 11,  119903]
  [arXiv:1309.2030 [hep-ex]].

\bibitem{Carena:2007ua}
  M.~S.~Carena, E.~Ponton, J.~Santiago and C.~E.~M.~Wagner,
  ``Electroweak constraints on warped models with custodial symmetry,''
  Phys.\ Rev.\ D {\bf 76} (2007) 035006
  [hep-ph/0701055].

\bibitem{Anastasiou:2009rv}
  C.~Anastasiou, E.~Furlan and J.~Santiago,
  ``Realistic Composite Higgs Models,''
  Phys.\ Rev.\ D {\bf 79} (2009) 075003
  [arXiv:0901.2117 [hep-ph]].

\bibitem{Grojean:2013qca}
  C.~Grojean, O.~Matsedonskyi and G.~Panico,
  ``Light top partners and precision physics,''
  JHEP {\bf 1310} (2013) 160
  [arXiv:1306.4655 [hep-ph]].

\bibitem{BiancofioreDaRoldJagerGPWeilertoappear}
P. Biancofiore, L. Da Rold, S. Jager, G. Perez and A. Weiler,
\textit{to appear}.

\bibitem{Schmaltz:2002wx}
  M.~Schmaltz,
  ``Physics beyond the standard model (theory): Introducing the little Higgs,''
  Nucl.\ Phys.\ Proc.\ Suppl.\  {\bf 117} (2003) 40
  [hep-ph/0210415].

\bibitem{Ciuchini:2013pca}
  M.~Ciuchini, E.~Franco, S.~Mishima and L.~Silvestrini,
  ``Electroweak Precision Observables, New Physics and the Nature of a 126 GeV Higgs Boson,''
  JHEP {\bf 1308} (2013) 106
  [arXiv:1306.4644 [hep-ph]].

\bibitem{Matsedonskyi:2012ym}
  O.~Matsedonskyi, G.~Panico and A.~Wulzer,
  ``Light Top Partners for a Light Composite Higgs,''
  JHEP {\bf 1301} (2013) 164
  [arXiv:1204.6333 [hep-ph]].

\bibitem{Redi:2012ha}
  M.~Redi and A.~Tesi,
  ``Implications of a Light Higgs in Composite Models,''
  JHEP {\bf 1210} (2012) 166
  [arXiv:1205.0232 [hep-ph]].

\bibitem{Marzocca:2012zn}
  D.~Marzocca, M.~Serone and J.~Shu,
  ``General Composite Higgs Models,''
  JHEP {\bf 1208} (2012) 013
  [arXiv:1205.0770 [hep-ph]].

\bibitem{Pomarol:2012qf}
  A.~Pomarol and F.~Riva,
  ``The Composite Higgs and Light Resonance Connection,''
  JHEP {\bf 1208} (2012) 135
  [arXiv:1205.6434 [hep-ph]].

\bibitem{Panico:2012uw}
  G.~Panico, M.~Redi, A.~Tesi and A.~Wulzer,
  ``On the Tuning and the Mass of the Composite Higgs,''
  JHEP {\bf 1303} (2013) 051
  [arXiv:1210.7114 [hep-ph]].

\bibitem{Pappadopulo:2013vca}
  D.~Pappadopulo, A.~Thamm and R.~Torre,
  ``A minimally tuned composite Higgs model from an extra
  JHEP {\bf 1307} (2013) 058
  [arXiv:1303.3062 [hep-ph]].

\bibitem{Barnard:2013hka}
  J.~Barnard, T.~Gherghetta, A.~Medina and T.~S.~Ray,
  ``Radiative corrections to the composite Higgs mass from a gluon partner,''
  JHEP {\bf 1310} (2013) 055
  [arXiv:1307.4778 [hep-ph]].

\bibitem{Carmona:2014iwa}
  A.~Carmona and F.~Goertz,
  ``A naturally light Higgs without light Top Partners,''
  arXiv:1410.8555 [hep-ph].

\bibitem{Randall:1999ee}
  L.~Randall and R.~Sundrum,
  ``A Large mass hierarchy from a small extra dimension,''
  Phys.\ Rev.\ Lett.\  {\bf 83} (1999) 3370
  [hep-ph/9905221].

\bibitem{ArkaniHamed:2000ds}
  N.~Arkani-Hamed, M.~Porrati and L.~Randall,
  ``Holography and phenomenology,''
  JHEP {\bf 0108} (2001) 017
  [hep-th/0012148].

\bibitem{Agashe:2006hk}
  K.~Agashe, A.~Belyaev, T.~Krupovnickas, G.~Perez and J.~Virzi,
  ``LHC Signals from Warped Extra Dimensions,''
  Phys.\ Rev.\ D {\bf 77} (2008) 015003
  [hep-ph/0612015].

\bibitem{Lillie:2007yh}
  B.~Lillie, L.~Randall and L.~T.~Wang,
  ``The Bulk RS KK-gluon at the LHC,''
  JHEP {\bf 0709} (2007) 074
  [hep-ph/0701166].

\bibitem{Agashe:2007ki}
  K.~Agashe, H.~Davoudiasl, S.~Gopalakrishna, T.~Han, G.~Y.~Huang, G.~Perez, Z.~G.~Si and A.~Soni,
  ``LHC Signals for Warped Electroweak Neutral Gauge Bosons,''
  Phys.\ Rev.\ D {\bf 76} (2007) 115015
  [arXiv:0709.0007 [hep-ph]].

\bibitem{Vignaroli:2014bpa}
  N.~Vignaroli,
  ``New W-prime signals at the LHC,''
  Phys.\ Rev.\ D {\bf 89} (2014) 095027
  [arXiv:1404.5558 [hep-ph]].

\bibitem{Carena:2007tn}
  M.~Carena, A.~D.~Medina, B.~Panes, N.~R.~Shah and C.~E.~M.~Wagner,
  ``Collider phenomenology of gauge-Higgs unification scenarios in warped extra dimensions,''
  Phys.\ Rev.\ D {\bf 77} (2008) 076003
  [arXiv:0712.0095 [hep-ph]].

\bibitem{SantiagoTalk}
J. Santiago,
\newblock {Talks at the Workshop on High Precision LHC Physics, May
  2014 and 
  Topical workshop on top quark differential distributions, 
  Sept 2014.}   

\bibitem{Barcelo:2011vk}
  R.~Barcelo, A.~Carmona, M.~Masip and J.~Santiago,
  ``Stealth gluons at hadron colliders,''
  Phys.\ Lett.\ B {\bf 707} (2012) 88
  [arXiv:1106.4054 [hep-ph]].

\bibitem{Barcelo:2011wu}
  R.~Barcelo, A.~Carmona, M.~Chala, M.~Masip and J.~Santiago,
  ``Single Vectorlike Quark Production at the LHC,''
  Nucl.\ Phys.\ B {\bf 857} (2012) 172
  [arXiv:1110.5914 [hep-ph]].

\bibitem{Bini:2011zb}
  C.~Bini, R.~Contino and N.~Vignaroli,
  ``Heavy-light decay topologies as a new strategy to discover a heavy gluon,''
  JHEP {\bf 1201} (2012) 157
  [arXiv:1110.6058 [hep-ph]].

\bibitem{Carmona:2012jk}
  A.~Carmona, M.~Chala and J.~Santiago,
  JHEP {\bf 1207} (2012) 049
  [arXiv:1205.2378 [hep-ph]].

\bibitem{Chala:2013ega}
  M.~Chala and J.~Santiago,
  ``$Hb\bar{b}$ production in composite Higgs models,''
  Phys.\ Rev.\ D {\bf 88} (2013) 3,  035010
  [arXiv:1305.1940 [hep-ph]].

\bibitem{Agashe:2004rs}
  K.~Agashe, R.~Contino and A.~Pomarol,
  ``The Minimal composite Higgs model,''
  Nucl.\ Phys.\ B {\bf 719} (2005) 165
  [hep-ph/0412089].

\bibitem{Contino:2006qr}
  R.~Contino, L.~Da Rold and A.~Pomarol,
  ``Light custodians in natural composite Higgs models,''
  Phys.\ Rev.\ D {\bf 75} (2007) 055014
  [hep-ph/0612048].

\bibitem{Agashe:2006at}
  K.~Agashe, R.~Contino, L.~Da Rold and A.~Pomarol,
  ``A Custodial symmetry for $Z b \bar{b}$,''
  Phys.\ Lett.\ B {\bf 641} (2006) 62
  [hep-ph/0605341].

\bibitem{DeSimone:2012fs}
  A.~De Simone, O.~Matsedonskyi, R.~Rattazzi and A.~Wulzer,
  JHEP {\bf 1304} (2013) 004
  [arXiv:1211.5663 [hep-ph]].

\bibitem{Petteni:2011np}
  M.~Petteni [ATLAS Collaboration],
  ``A Search for $t\bar{t}$ Resonances in the Dilepton Channel in 1.04 fb$^{-1}$ of $pp$ collisions at $\sqrt{s}=7$ TeV with the ATLAS experiment,''
  arXiv:1111.6933 [hep-ex].

\bibitem{CMS:2014rna}
  CMS Collaboration [CMS Collaboration],
  ``Search for ttbar resonances in dilepton+jets final states in pp collisions at 8 TeV,''
  CMS-PAS-B2G-12-007.

\bibitem{CMS:2013vca}
  CMS Collaboration [CMS Collaboration],
  ``Search for Anomalous Top Quark Pair Production in the Boosted All-Hadronic Final State using pp Collisions at sqrt(s) = 8 TeV,''
  CMS-PAS-B2G-12-005.

\bibitem{Aad:2012raa}
  G.~Aad {\it et al.}  [ATLAS Collaboration],
  ``Search for resonances decaying into top-quark pairs using fully hadronic decays in $pp$ collisions with ATLAS at $\sqrt{s}=7$ TeV,''
  JHEP {\bf 1301} (2013) 116
  [arXiv:1211.2202 [hep-ex]].

\bibitem{Chatrchyan:2013wfa}
  S.~Chatrchyan {\it et al.}  [CMS Collaboration],
  ``Search for top-quark partners with charge 5/3 in the same-sign dilepton final state,''
  Phys.\ Rev.\ Lett.\  {\bf 112} (2014) 171801
  [arXiv:1312.2391 [hep-ex]].

\bibitem{Aguilar-Saavedra:2013qpa}
  J.~A.~Aguilar-Saavedra, R.~Benbrik, S.~Heinemeyer and M.~Pérez-Victoria,
  ``Handbook of vectorlike quarks: Mixing and single production,''
  Phys.\ Rev.\ D {\bf 88} (2013) 9,  094010
  [arXiv:1306.0572 [hep-ph]].

\bibitem{Matsedonskyi:2014mna}
  O.~Matsedonskyi, G.~Panico and A.~Wulzer,
  ``On the Interpretation of Top Partners Searches,''
  arXiv:1409.0100 [hep-ph].

\bibitem{Alloul:2013bka}
  A.~Alloul, N.~D.~Christensen, C.~Degrande, C.~Duhr and B.~Fuks,
  ``FeynRules  2.0 - A complete toolbox for tree-level phenomenology,''
  Comput.\ Phys.\ Commun.\  {\bf 185} (2014) 2250
  [arXiv:1310.1921 [hep-ph]].

\bibitem{Maltoni:2002qb}
  F.~Maltoni and T.~Stelzer,
  ``MadEvent: Automatic event generation with MadGraph,''
  JHEP {\bf 0302} (2003) 027
  [hep-ph/0208156].

\bibitem{Sjostrand:2006za}
  T.~Sjostrand, S.~Mrenna and P.~Z.~Skands,
  ``PYTHIA 6.4 Physics and Manual,''
  JHEP {\bf 0605} (2006) 026
  [hep-ph/0603175].

\bibitem{Mangano:2006rw}
  M.~L.~Mangano, M.~Moretti, F.~Piccinini and M.~Treccani,
  ``Matching matrix elements and shower evolution for top-quark production in hadronic collisions,''
  JHEP {\bf 0701} (2007) 013
  [hep-ph/0611129].

\bibitem{Nadolsky:2008zw}
  P.~M.~Nadolsky, H.~L.~Lai, Q.~H.~Cao, J.~Huston, J.~Pumplin, D.~Stump, W.~K.~Tung and C.-P.~Yuan,
  ``Implications of CTEQ global analysis for collider observables,''
  Phys.\ Rev.\ D {\bf 78} (2008) 013004
  [arXiv:0802.0007 [hep-ph]].

\bibitem{Czakon:2013goa}
  M.~Czakon, P.~Fiedler and A.~Mitov,
  ``Total Top-Quark Pair-Production Cross Section at Hadron Colliders Through $\mathcal{O}(\alpha^4_S)$,''
  Phys.\ Rev.\ Lett.\  {\bf 110} (2013) 252004
  [arXiv:1303.6254 [hep-ph]].

\bibitem{Cacciari:2011ma}
  M.~Cacciari, G.~P.~Salam and G.~Soyez,
  ``FastJet User Manual,''
  Eur.\ Phys.\ J.\ C {\bf 72} (2012) 1896
  [arXiv:1111.6097 [hep-ph]].

\bibitem{Cacciari:2008gp}
  M.~Cacciari, G.~P.~Salam and G.~Soyez,
  ``The Anti-k(t) jet clustering algorithm,''
  JHEP {\bf 0804} (2008) 063
  [arXiv:0802.1189 [hep-ph]].

\bibitem{Aad:2013nca}
  G.~Aad {\it et al.}  [ATLAS Collaboration],
  ``Search for $t\bar t$ resonances in the lepton plus jets final state with ATLAS using 4.7 fb$^{-1}$ of $pp$ collisions at $\sqrt{s} = 7$ TeV,''
  Phys.\ Rev.\ D {\bf 88} (2013) 1,  012004
  [arXiv:1305.2756 [hep-ex]].

\bibitem{Kaplan:2008ie}
  D.~E.~Kaplan, K.~Rehermann, M.~D.~Schwartz and B.~Tweedie,
  ``Top Tagging: A Method for Identifying Boosted Hadronically Decaying Top Quarks,''
  Phys.\ Rev.\ Lett.\  {\bf 101} (2008) 142001
  [arXiv:0806.0848 [hep-ph]].


\bibitem{Krohn:2009th}
  D.~Krohn, J.~Thaler and L.~T.~Wang,
  ``Jet Trimming,''
  JHEP {\bf 1002} (2010) 084
  [arXiv:0912.1342 [hep-ph]].

\bibitem{ATL-PHYS-PUB-2013-003}
CERN preprint ATL-PHYS-PUB-2013-003 (2013).

\bibitem{Backovic:2014uma}
  M.~Backović, G.~Perez, T.~Flacke and S.~J.~Lee,
  ``LHC Top Partner Searches Beyond the 2 TeV Mass Region,''
  arXiv:1409.0409 [hep-ph].

\bibitem{Almeida:2010pa}
  L.~G.~Almeida, S.~J.~Lee, G.~Perez, G.~Sterman and I.~Sung,
  ``Template Overlap Method for Massive Jets,''
  Phys.\ Rev.\ D {\bf 82} (2010) 054034
  [arXiv:1006.2035 [hep-ph]].

\bibitem{Backovic:2013bga}
  M.~Backovic, O.~Gabizon, J.~Juknevich, G.~Perez and Y.~Soreq,
  ``Measuring boosted tops in semi-leptonic $t\bar t$ events for the standard model and beyond,''
  JHEP {\bf 1404} (2014) 176
  [arXiv:1311.2962 [hep-ph]].

\bibitem{WillocqTalk}
S. Willocq,
\newblock {Talk at the ECFA High Luminosity LHC Experiments 
Workshop - Oct 2014}.

\bibitem{Greco:2014aza}
  D.~Greco and D.~Liu,
  arXiv:1410.2883 [hep-ph].

\end{thebibliography}

\end{document}